\documentclass[%
 aip,
 amsmath,amssymb,
 reprint,%
]{revtex4-1}

\usepackage{graphicx}% Include figure files
\usepackage{dcolumn}% Align table columns on decimal point
\usepackage{bm}% bold math
\usepackage{caption}
\usepackage{subfigure}
\usepackage{amssymb}
\usepackage{multirow}
\usepackage{makecell}
\usepackage[utf8]{inputenc}
\usepackage[T1]{fontenc}
\usepackage{float}

\usepackage{hyperref}
\hypersetup{
    colorlinks,    citecolor=blue,    filecolor=blue,    linkcolor=blue,    urlcolor=blue
}
\usepackage{soul}
\setstcolor{red}
\usepackage{siunitx}
\begin{document}

% \preprint{AIP/123-QED}

% \title{Penetration of a relativistic plasma-generated ionization wave into a strong magnetic field}
\title{Emission of electromagnetic waves as a stopping mechanism for nonlinear collisionless ionization waves in a high-$\beta$ regime}

%%\title{Dynamics of a relativistic velocity collisionless ionization wave in an applied magnetic field}
% Force line breaks with \\

\author{Haotian Mao}
\affiliation{Department of Mechanical and Aerospace Engineering, University of California at San Diego, La Jolla, California 92093, USA}
 %\altaffiliation{Mechanical \& Aerospace Department, University of California, San Diego.}%Lines break automatically or can be forced with \\
\author{Kathleen Weichman}
\affiliation{Department of Mechanical and Aerospace Engineering, University of California at San Diego, La Jolla, California 92093, USA}
\affiliation{University of Rochester, Laboratory for Laser Energetics, Rochester, New York 14623, USA}

\author{Zheng Gong}
\affiliation{Center for High Energy Density Science, University of Texas at Austin, Texas 78712, USA}
\affiliation{School of Physics, Peking University, 100871, People’s Republic of China}

\author{Todd Ditmire}
\author{Hernan Quevedo}
\affiliation{Center for High Energy Density Science, University of Texas at Austin, Texas 78712, USA}

\author{Alexey Arefiev}%
 %\email{Second.Author@institution.edu.}
\affiliation{Department of Mechanical and Aerospace Engineering, University of California at San Diego, La Jolla, California 92093, USA}
\affiliation{Center for Energy Research, University of California at San Diego, La Jolla, California 92037, USA}

\date{\today}% It is always \today, today,

\begin{abstract}

A high energy density plasma embedded in a neutral gas is able to launch an outward-propagating nonlinear electrostatic ionization wave that traps energetic electrons. The trapping maintains a strong sheath electric field, enabling rapid and long-lasting wave propagation aided by field ionization. Using 1D3V kinetic simulations, we examine the propagation of the ionization wave in the presence of a transverse MG-level magnetic field with the objective to identify qualitative changes in a regime where the initial thermal pressure of the plasma exceeds the pressure of the magnetic field ($\beta>1$). Our key finding is that the magnetic field stops the propagation by causing the energetic electrons sustaining the wave to lose their energy by emitting an electromagnetic wave. The emission is accompanied by the magnetic field expulsion from the plasma and an increased electron loss from the trapping wave structure. The described effect provides a mechanism mitigating rapid plasma expansion for those applications that involve an embedded plasma, such as high-flux neutron production from laser-irradiated deuterium gas jets. 

\end{abstract}

\maketitle

%+++++++++++++++++++++++++++++++++++++++++++++++++++++++++++++++++
\section{\label{sec:level1}Introduction}

The development of high-power lasers~\cite{danson2019lasers} has enabled the generation of high energy density (HED) plasmas in laboratory conditions. Such plasmas are essential to various applications, including inertial confinement fusion~\cite{lindl1995icf}, laboratory astrophysics~\cite{bulanov2015labastro}, and energetic particle~\cite{macchi2013tnsa_review,pomerantz2014neutrons} and radiation~\cite{kmetec1992brem,stark2016xray} sources.
The addition of strong magnetic fields (applied or plasma-generated) enables a host of novel magnetization-related phenomena, for example in the areas of magnetic reconnection~\cite{fiskel2014reconnection}, direct laser acceleration~\cite{arefiev2015electrons,arefiev2020dla}, hot electron transport~\cite{johzaki2015transport,bailly2018guiding}, and ion acceleration~\cite{arefiev2016protons,weichman2020protons,kuri2018rpa_cp,cheng2019rpa}.

Laboratory laser-generated HED plasmas are often highly localized, so their expansion plays an important role in plasma dynamics and energy and particle transport. Plasma expansion is an extensively studied topic that continues to attract interest due to the complexity and diversity of the phenomena associated with it. For example, inward plasma expansion that can be achieved by using targets with inner cavities~\cite{murakami.2018} differs from the conventional and well-studied outwards plasma expansion into vacuum~\cite{gurevich1966expansion,mora2003expansion}. The novel features of the inward expansion are short flashes of energetic protons~\cite{murakami.2019} and generation of extreme electric~\cite{koga.2020} and magnetic fields~\cite{murakami.2020,kathleen2020Bfieldamplify}.

%%For example, outward expansion of a plasma slab into vacuum~\cite{gurevich1966expansion,mora2003expansion} is one of primary mechanisms for energy transfer from hot electrons to ions~\cite{wilks.2001}. It is employed extensively in experiments to generate energetic ion beams for proton radiography~\cite{Zystra.2012}. However, the change from outward to inward expansion, which can be achieved by using targets with inner cavities, has been recently shown to introduce new physics regimes that include flashes of energetic protons and generation of extreme electric and magnetic fields. 

There are applications where the generated HED plasma is surrounded by un-ionized gas rather than by vacuum. One example is the production of monoenergetic neutrons with high flux from a laser-irradiated clustering deuterium gas jet~\cite{WBang2013,WBangPOP2013,WBangPRE2013}. In this case, the laser generates within the gas jet an embedded plasma filament with energetic electrons and ions. The neutrons are produced via DD fusion reactions. The reaction rate decreases as the plasma expands and the density of energetic ions drops, so the rate of plasma expansion into ambient gas plays an important role for this application.

A recent experimental campaign aimed at understanding plasma expansion into ambient gas has revealed that the evolution of an HED plasma filament surrounded by gas differs from that of a plasma filament in vacuum~\cite{mccormick2014soliton}. The plasma boundary expands with a velocity greatly exceeding that of the ions. The expansion is associated with rapid plasma production through collisionless field ionization whose rate is much higher than the rate of impact ionization. The underlying mechanism is a unique nonlinear electrostatic ionization wave. The sheath electric field of the HED plasma ionizes the gas, which allows the plasma boundary to expand on a time scale independent of the ion dynamics. The electrons produced via the field ionization move into the plasma and generate an electric field whose sign is opposite to that of the sheath field. This structure traps energetic electrons next to the plasma boundary. As the boundary expands, the energetic electrons remain bunched and maintain a strong sheath field. The described ionization wave has been shown to propagate with a relativistic velocity over distances that greatly exceed the spatial scale of the sheath~\cite{mccormick2014soliton}. 

The focus of this paper is on the propagation of the ionization wave, launched by an HED plasma, through ambient gas with a strong applied magnetic field. The effect of magnetic fields on plasma expansion into vacuum has been extensively studied~\cite{katz1961boundary,anderson1980magnetic,ditmire2000expansion,rubio2016magnetic}. This computational work is the first study of the magnetic field impact on the ionization wave mediated plasma expansion into neutral gas. We consider a regime where the initial thermal pressure of the plasma exceeds the pressure of the applied magnetic field ($\beta > 1$). Our key finding is that the magnetic field stops the propagation of the ionization wave by causing the energetic electrons sustaining the wave to lose their energy by emitting an electromagnetic wave. We find that the emission is accompanied by the magnetic field expulsion from the plasma and an increased electron loss from the trapping wave structure located next to the plasma boundary.

Our choice of the plasma parameters for this study is informed by the experiments where the ionization waves have been previously observed~\cite{mccormick2014soliton}. Specifically, the initial characteristic electron energy is several hundred keV and the electron density is $\sim 3 \times 10^{19}$~cm$^{-3}$. The applied magnetic fields are in the MG range. Such strong fields can be generated using various experimental techniques~\cite{portugall1997field_gen,portugall1999field_gen,fujioka2013coil,santos2015coil,santos2018coil,gao2016coil,goyon2017coil,ivanov2018zebra}. For the purpose of this study, the applied field is treated as static, since our time-scale of interest ($\sim 10$~ps) is much shorter than the life-time of the experimentally generated fields.

Our results provide insights for an experimental setup where an HED plasma is embedded into neutral gas in the presence of an applied magnetic field. Such a setup is being considered for the purpose of boosting the neutron yield from laser-irradiated clustering deuterium gas jets~\cite{WBang2013,WBangPOP2013,WBangPRE2013}. The expectation is that the magnetic field would slow down the transverse plasma expansion and delay the decrease of the ion density that reduces the rate of DD fusion reactions. The rapid expansion of the plasma boundary due to the ionization wave and the associated with it expulsion of the magnetic field would negate the impact of the magnetic field on the energetic ions within the plasma. Therefore, the stopping of the ionization wave by the magnetic field discussed in this work can be viewed as a mitigating mechanism.

The outline of this paper is as follows. In Section~\ref{sec:non-mag}, we highlight main properties of the ionization wave propagation in the absence of a magnetic field. In Section~\ref{sec:estiamtes}, we present qualitative estimates for the wave propagation through an applied magnetic field in the high-$\beta$ regime and the associated emission of an electromagnetic wave. In Section~\ref{sec:2MG}, we examine results of a kinetic simulations for a 2~MG field and compare them to the estimates from Sec.~\ref{sec:estiamtes}. In Section~\ref{sec:stopping},  we present results of a $\beta$-scan, finding the stopping distance for the ionization wave and the fraction of the wave's energy emitted by the electromagnetic wave. In Section~\ref{sec:summary}, we summarize our findings and discuss potential extensions to this work.

\section{Ionization wave without applied magnetic field} \label{sec:non-mag}

In this section, we review the salient features of the ionization wave and its propagation through ambient gas in the absence of an applied magnetic field. The goal is to provide the necessary context for the study of the ionization wave propagation with an applied field presented in Sections~\ref{sec:estiamtes}, \ref{sec:2MG}  and~\ref{sec:stopping}.

%%The expansion of a high energy density plasma filament into a surrounding neutral gas differs substantially from the expansion of a hot plasma into vacuum. While we will discuss the effect a magnetic field can have on restricting this expansion in Sections~\ref{sec:ramp} and~\ref{sec:constant}, in this Section we will first summarize the salient details of the expansion process in a non-magnetized case.

The ionization wave is a collisionless mechanism by which the boundary of an HED plasma embedded in a neutral gas rapidly expands over a prolonged time period without involving the ion dynamics. Reference~[\onlinecite{mccormick2014soliton}] provides experimental results confirming its existence. The wave has been shown to exist for a cylindrical and a slab-like plasma, as demonstrated by one and two-dimensional kinetic simulations~\cite{mccormick2014soliton}. In order to simplify our analysis while retaining the key physics, we consider a one-dimensional case in this paper.

\begin{figure}
\centering
\begin{minipage}[b]{0.391\textwidth}
\includegraphics[width=1\textwidth]{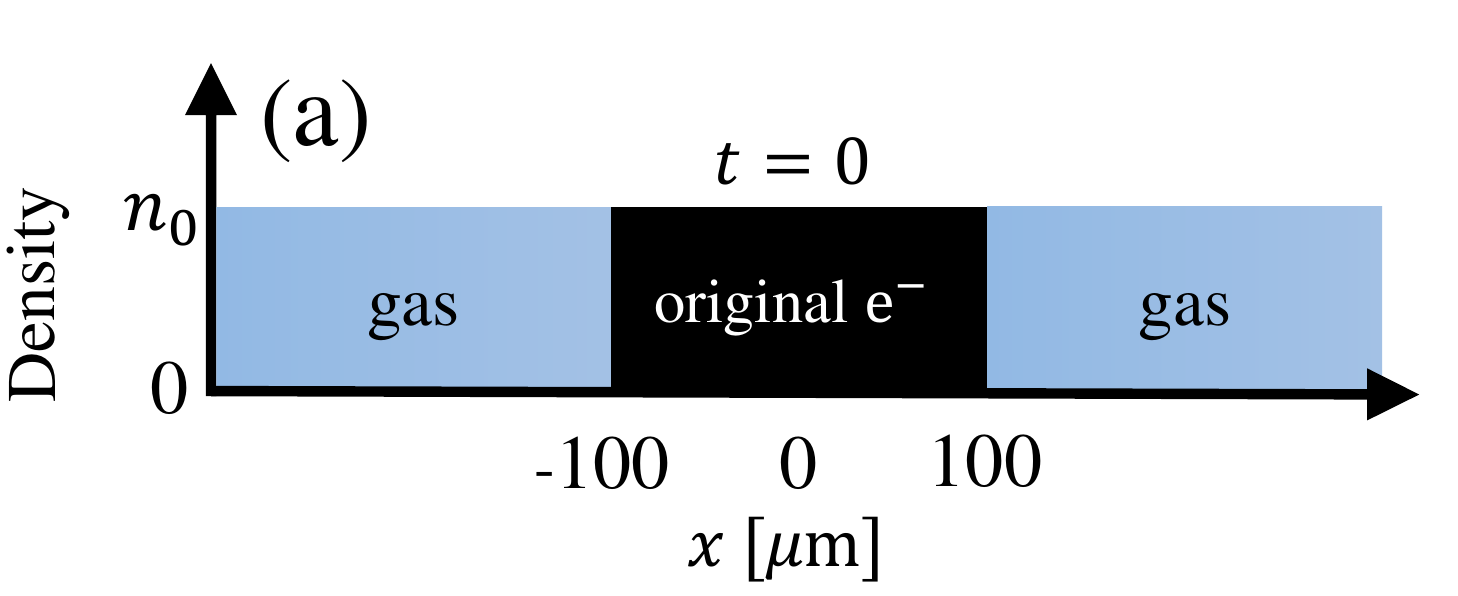}
\end{minipage}
\begin{minipage}[b]{0.391\textwidth}
\includegraphics[width=1\textwidth]{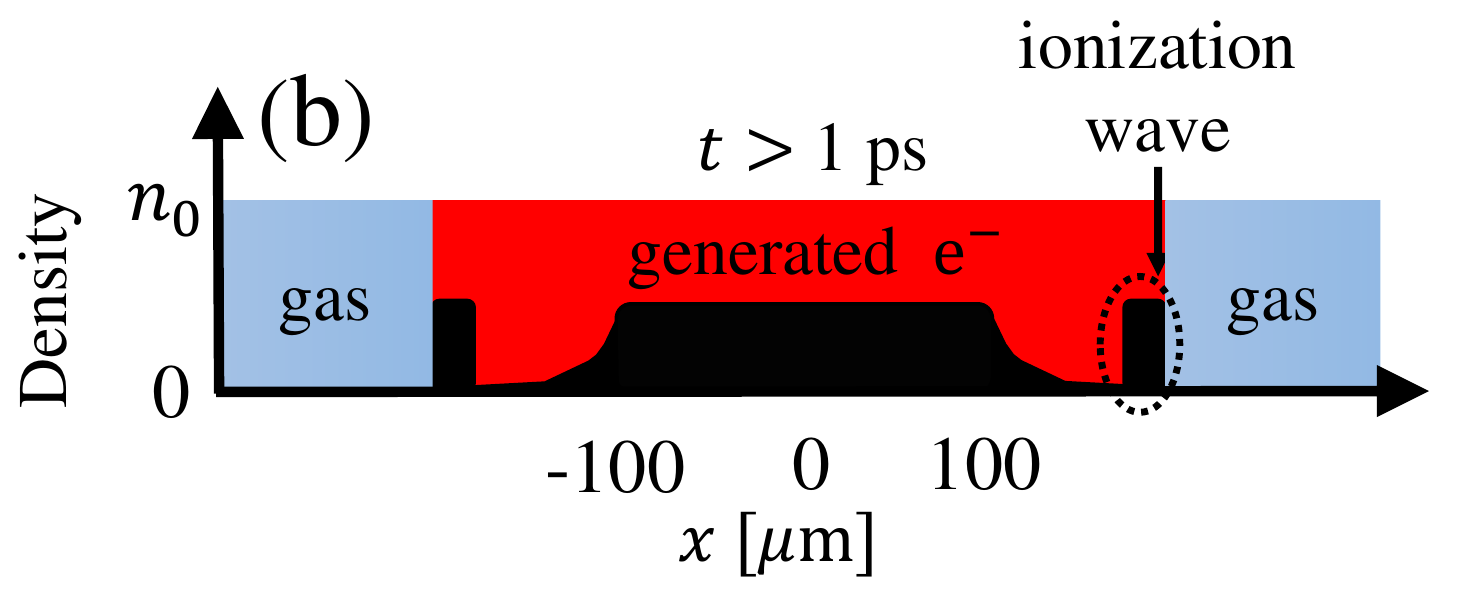}
\end{minipage}
\caption{Setup for 1D PIC simulations (a) and a schematic structure of this system with a propagating ionization wave (b). The blue color represents neutral hydrogen gas, the black color represents the original electrons initialized at $t=0$ (original electrons), and the red color represent the electrons generated through field ionization of the gas (generated electrons).}
%%Schematic of 1D particle-in-cell (PIC) simulation of the ionization wave formation. (a)~The initial set up of the PIC simulation. The black and light blue area respectively denotes the electrons originally from the plasma (original electrons) and the ambient hydrogen gas. (b)~A schematic of ionization wave formation. The black area denotes the original electrons and the red area denotes the electrons born from ionization (generated electrons). The original electrons forms a structure which detaches from the plasma at $t>1$~ps. We refer to this structure as the ionization wave.
\label{fig1}
\end{figure}

The setup that we use to study the ionization wave with the help of particle-in-cell (PIC) simulations is shown in Fig.~\ref{fig1}(a). A hydrogen plasma slab with hot electrons is embedded into a neutral hydrogen gas. Initially, the plasma electrons have a spatially uniform water-bag momentum distribution with $|p_x| \leq m_e c$, where $m_e$ is the electron mass and $c$ is the speed of light. There is no transverse momentum. The width of the plasma slab is 200~$\mu$m ($|x|<100$~$\mu$m) and it is sufficiently large for the formation of the ionization wave not to depend on its size. The plasma density of $3.3 \times 10^{19}$~cm$^{-3}$ is chosen to be representative of high energy density experiments where ionization waves have been observed in the past~\cite{mccormick2014soliton}. The density of the plasma is equal to the density of the ambient gas. %Additional
Detailed simulation parameters are given in Table~\ref{simulation}. 

We model the plasma dynamics using Cartesian 1D3V PIC simulations with the open source PIC code EPOCH~\cite{arber2015epoch}. The primary ionization mechanism for the ionization wave is the collisionless field ionization. We thus neglect collisional impact ionization in our simulations (see Ref.~[\onlinecite{mccormick2014soliton}] for a comparison of time-scales between the collisional and collisionless ionization mechanisms). The field ionization is modeled using a Monte Carlo ionization module that employs probabilistic ionization rates for tunneling and barrier suppression ionization mechanisms. The ionization module is discussed in more detail in Ref.~[\onlinecite{arber2015epoch}].

\begin{table}
      %\caption{}
      \centering
        \begin{tabular}{|c|c|}
        \hline
        \multicolumn{2}{|c|}{\bfseries Plasma Parameters}\\
        \hline
        \multicolumn{2}{|c|}{Fully ionized hydrogen plasma}\\
        \hline
        Plasma density $n_0$&  $3.3 \times 10^{19}$~cm$^{-3}$\\
        \hline
        Plasma location& $|x|< 100$~$\mu$m\\
        \hline
        \makecell{Electron momentum \\ distribution  \\($|x|<100$~$\mu$m)}& \makecell{Water-bag\\ (cutoff : \\$p_x = \pm m_ec$)}\\
        \hline
        \multicolumn{2}{|c|}{\bfseries Hydrogen gas Parameters}\\
        \hline
        Hydrogen density&$3.3 \times 10^{19}$~cm$^{-3}$\\
        \hline
        Hydrogen location& $|x| > 100$~$\mu$m\\
        \hline
        \multicolumn{2}{|c|}{\bfseries Simulation Parameters}\\
        \hline
        Simulation domain & [-600~$\mu$m , 600~$\mu$m]\\
        \hline
        Cells/micron& 100/$\mu$m\\
        \hline
        Particles/cell& 200\\
        \hline
        \multicolumn{2}{|c|}{\bfseries Boundary Condition}\\
        \hline
        Left boundary ($x= -600$~$\mu$m) & open\\
        \hline
        Right boundary ($x=600$~$\mu$m)& open\\
        \hline

        \end{tabular}
    \caption{Parameters for 1D PIC simulation. The plasma ions are initialized cold. The plasma is treated as collisionless. 
    % The ionization model used in this simulation is field ionization embedded in EPOCH.
    }
    \label{simulation}
\end{table}

\begin{figure}
\centering
\includegraphics[width=0.4\textwidth]{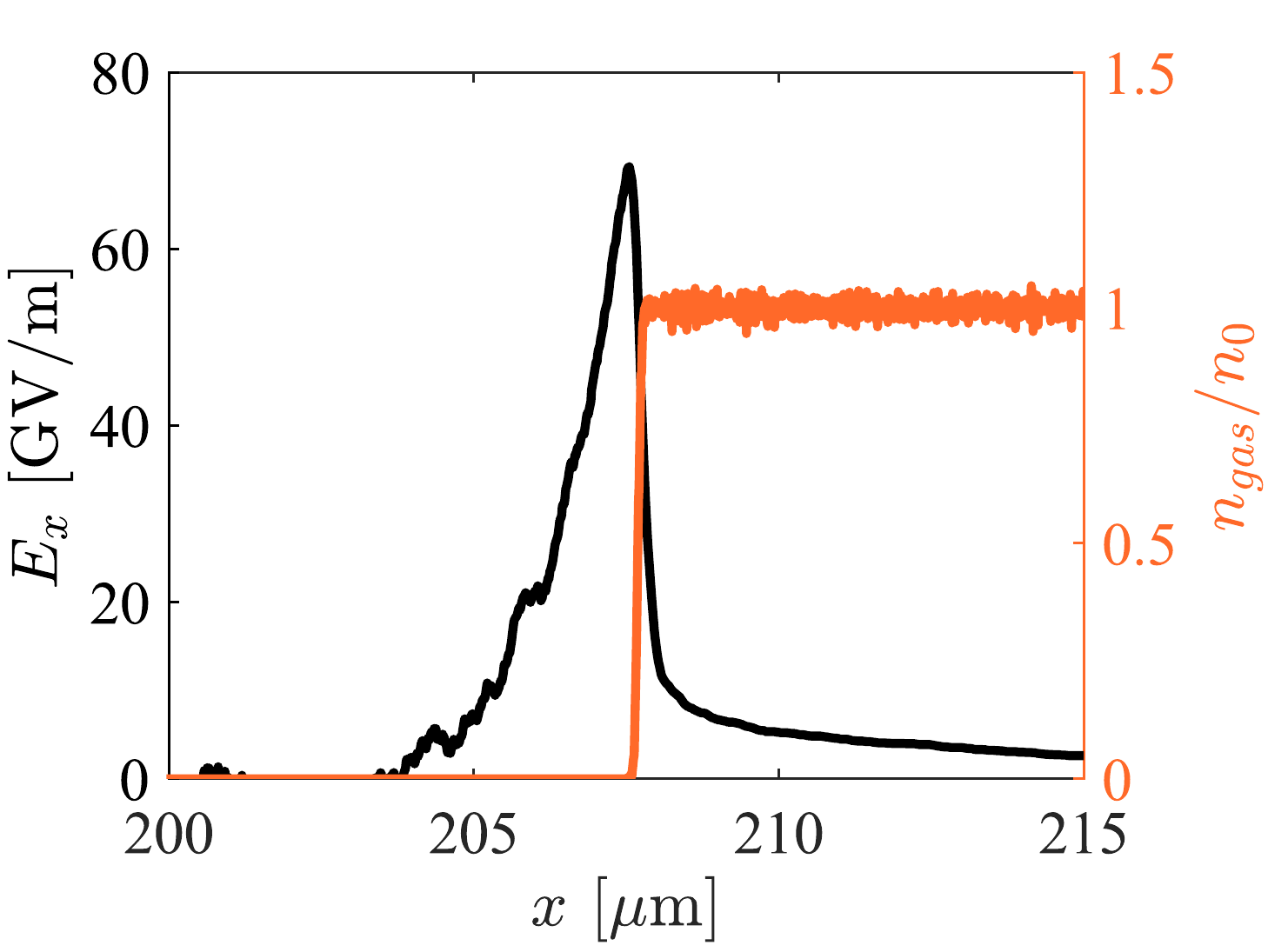}
\caption{Sheath electric field $E_x$ and neutral gas density $n_{gas}$ at the plasma edge expanding due to the ionization wave. The black curve is $E_x$ (left axis) and the orange curve is $n_{gas}$ (right axis) normalized to the initial density $n_0$. The snapshot is taken at $t=2$~ps.}
\label{fig-2-ver2}
\end{figure}

The hot electrons generate a sheath electric field at the edge of the ion density profile. This field prevents the electrons from freely streaming out into the surrounding gas. As shown in Fig.~\ref{fig-2-ver2}, this field, whose maximum strength is roughly $70$~GV/m, protrudes into the gas by roughly a Debye length. Due to the strength of the electric field, all atoms in the sheath become quickly ionized. The ionization effectively shifts the plasma boundary outwards and manifests itself as a plasma expansion. Note that the corresponding rate of expansion is independent of ion dynamics.

We distinguish the electrons produced as a result of the described ionization process from the `original electrons' and refer to them as `generated electrons'. The generated electrons have no kinetic energy when they appear at the plasma edge. They gain their kinetic energy from the sheath field as they are accelerated into the plasma. In the conventional picture of an expanding, homogeneous plasma slab, the sheath field confines both groups of electrons (generated and original) within the plasma where they are free to mix. The entire electron population can then be characterized by an average kinetic energy, $\varepsilon_{av}$. The strength of the sheath field is determined by $\varepsilon_{av}$. As the sheath field produces more and more generated electrons, the electron population cools down and $\varepsilon_{av}$ drops, which then reduces the strength of the field. The field ionization rate is sensitive to the field strength, so that is why one would expect for the described expansion to quickly terminate itself in the described scenario.

%%{The x-oriented sheath field and the density of ambient gas in the sheath region. The left vertical axis and blue line shows the sheath field strength. The right vertical axis and the orange line shows the normalized ambient hydrogen gas density $n_{H^2}$. The ambient gas interacts with the strong sheath field peak ($E_x\sim 10$~GV/m) inside region $l_{H^2}$, which is approximately the Debye length $\lambda_{De}$. The snapshot is taken at $t=2$~ps.}

\begin{figure}
\centering
\begin{minipage}[b]{0.45\textwidth}
\includegraphics[width=1\textwidth]{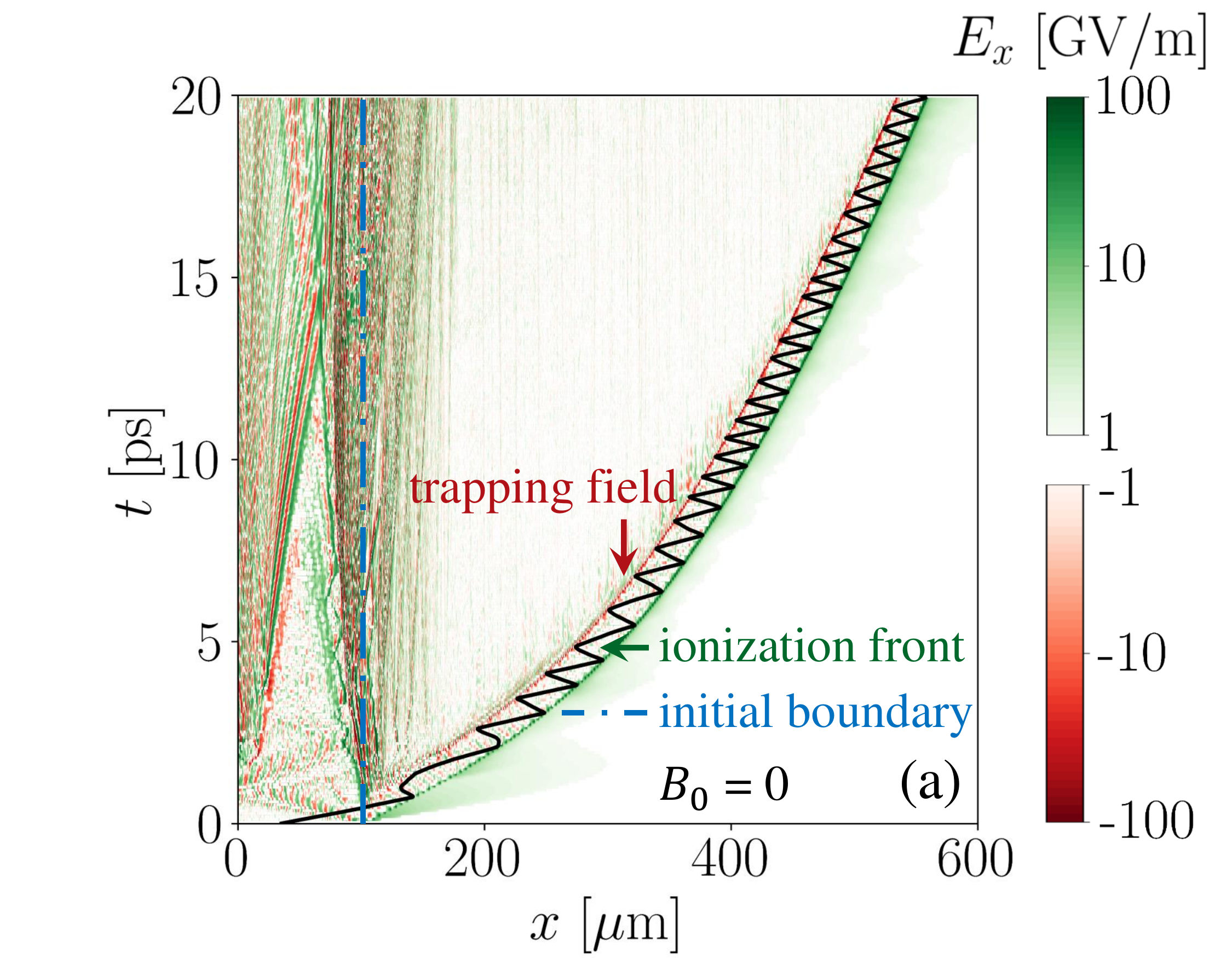}
\end{minipage}
\label{gradient-300(2)}
\begin{minipage}[b]{0.45\textwidth}
\includegraphics[width=1\textwidth]{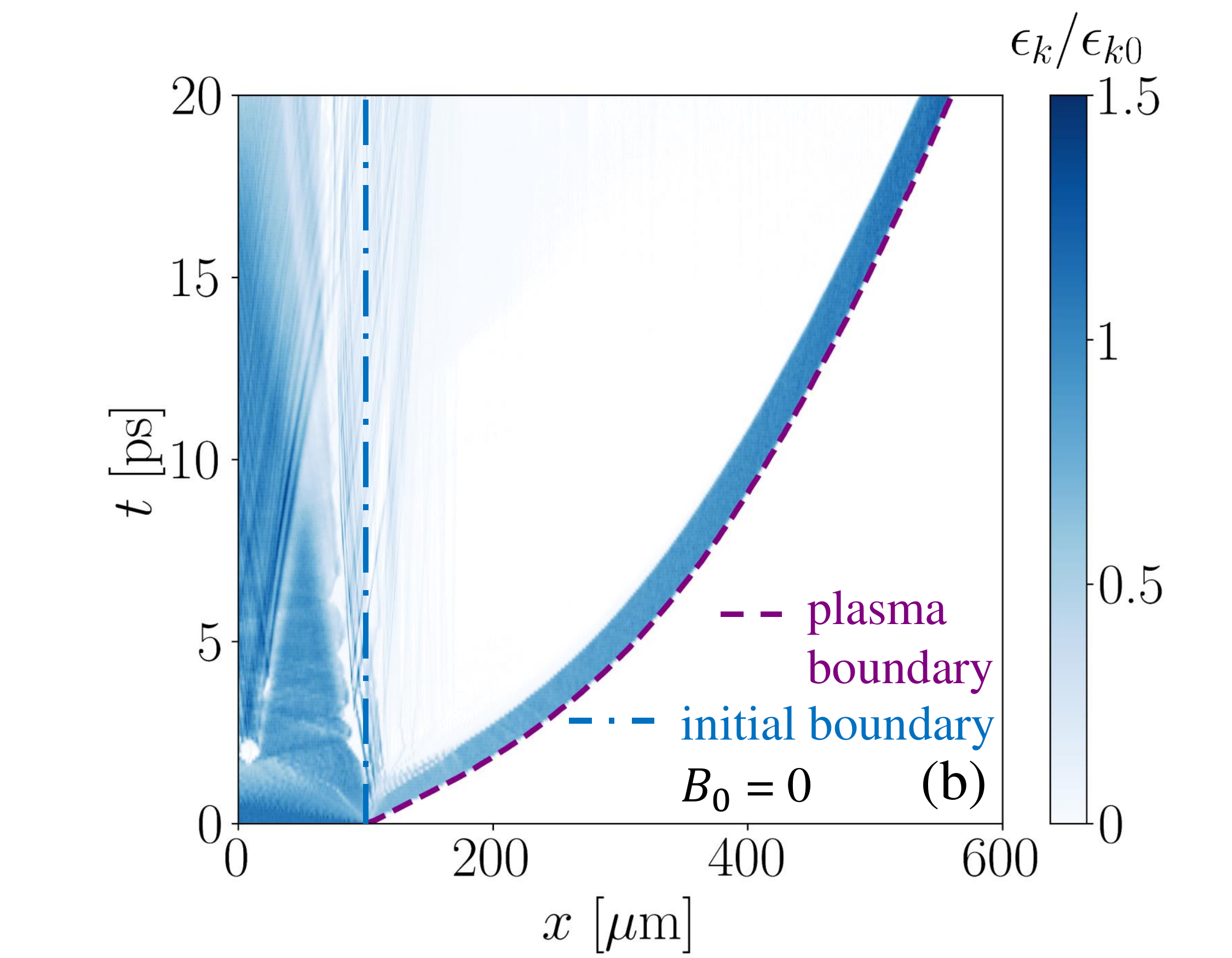}
\end{minipage}
\begin{minipage}[b]{0.45\textwidth}
\includegraphics[width=1\textwidth]{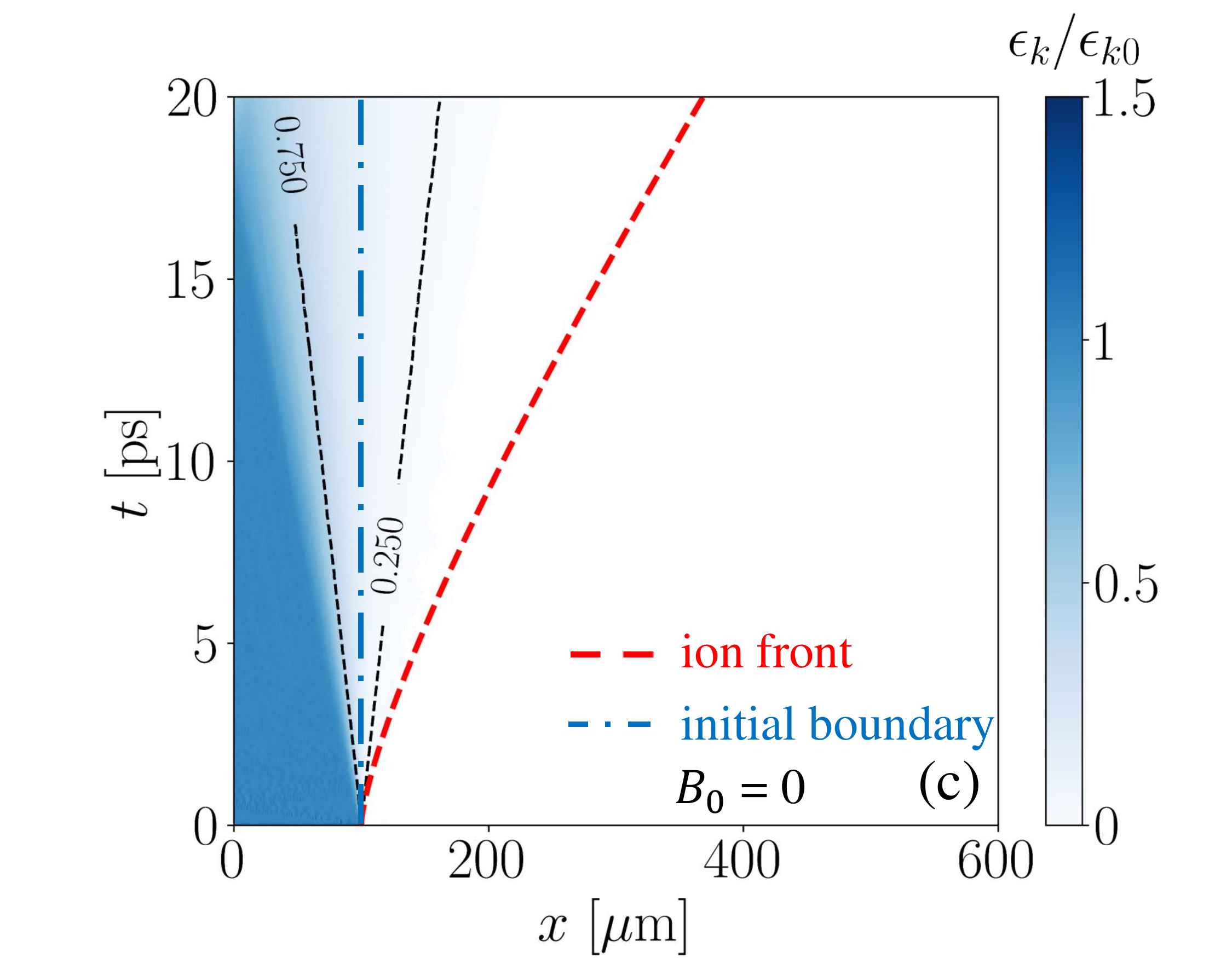}
\end{minipage}
\caption{Plasma expansion into neutral gas (a, b) and into vacuum (c). (a)~ Time evolution of $E_x$. The black curve shows a representative trajectory of a trapped energetic electron from the original population. (b, c)~Electron kinetic energy density $\epsilon_k$ normalized to its initial value $\epsilon_{k0}$. In panel (c), the black dashed lines are electron density contours ($n_e/n_0$) and the red dashed line is the ion front. }
\label{B-0}
\end{figure}

%\caption{The ionization wave structure without the applied magnetic field. (a)~Electron phase space. The grey dash-dotted line denotes the average momentum of the original electrons inside the soliton, which agrees well with the speed of the ionization wave ($\approx$ 0.15~$c$). (b)~Electric field in the soliton propagation ($x$) direction. (c)~Electron density, normalized by the initial electron density. Figures~(a)-(c) are snapshots at $t = 2$~ps.}

Figure~\ref{B-0}(b) shows that the plasma boundary in our setup continues to expand even after the plasma volume increases by more than a factor of five. Remarkably, the sheath electric field at the boundary remains strong during this expansion [see Fig.~\ref{B-0}(a)]. The prolonged propagation is enabled by electron trapping near the expanding boundary. A representative trajectory for a trapped energetic electron from the original population is shown in Fig.~\ref{B-0}(a). As shown in Fig.~\ref{B-0}(a), a negative electric field arises inside the plasma and travels together with the boundary expanding in the positive direction along the $x$-axis. Figures~\ref{fig1-ver2}(a) and \ref{fig1-ver2}(b) provide snapshots of the electron $(p_x, x)$ phase space and the electric field structure. The trapping field indeed keeps a group of energetic electrons bunched up next to the plasma boundary. The trapping maintains the sheath electric field at a nearly constant magnitude, enabling the long-lasting propagation.

The observed field structure is a nonlinear electrostatic wave in which the generated electrons create the trapping field for a group of original energetic electrons, as schematically shown in Fig.~\ref{fig1}(b). The wave consists of a sheath field spike and an adjacent trapping field spike with an opposite sign. The field structure can be viewed as a BGK wave sustained by field ionization, but an analytical solution is yet to be constructed for this mode. Figure~\ref{fig1-ver2}(c) shows the total electron density and the contributions from the generated and original electrons in the vicinity of the plasma boundary. The acceleration of the generated electrons into the plasma is evident from the drop in their density. The parent ions do not respond on this time scale, which leads to an excess of positive charge behind the  ionization front. The charge separation then creates a negative electric field inside the plasma that we refer to as the trapping field. It is important to note that most of the generated electrons escape the wave and only a group of the original electrons remains trapped. 

In the example illustrated in Figs.~\ref{B-0} and \ref{fig1-ver2}, the ionization wave travels with a relativistic speed for more than 20 ps. This significant propagation speed is also evident from the phase space plot in Fig.~\ref{fig1-ver2}(a) where the trapped electrons are shifted upwards. Over the course of the wave propagation, the field amplitude generated by the wave and the wave width remain relatively unchanged. The wave shows no sign of stopping at the end of our simulation. A distinctive feature of this expansion mechanism compared to the conventional plasma expansion into vacuum is that the ionization wave maintains a high level of energy density at the plasma boundary. Figures~\ref{B-0}(b) and \ref{B-0}(c) compare these two regimes of expansion by showing the evolution of the electron kinetic energy density, $\epsilon_k$, normalized to its initial value, $\epsilon_{k0}$.

\section{Qualitative analysis of wave propagation in a magnetic field} \label{sec:estiamtes}

In this section, we turn our attention to a setup where the ionization wave propagates through gas with an applied magnetic field. The magnetic field lines are parallel to the surface of the plasma. Without any loss of generality, we assume that the magnetic field is directed along the $z$-axis. Our goal is to qualitatively assess the impact of this field on the wave propagation in a high-$\beta$ regime.

\begin{figure}
\centering
\begin{minipage}[b]{0.391\textwidth}
\includegraphics[width=1\textwidth]{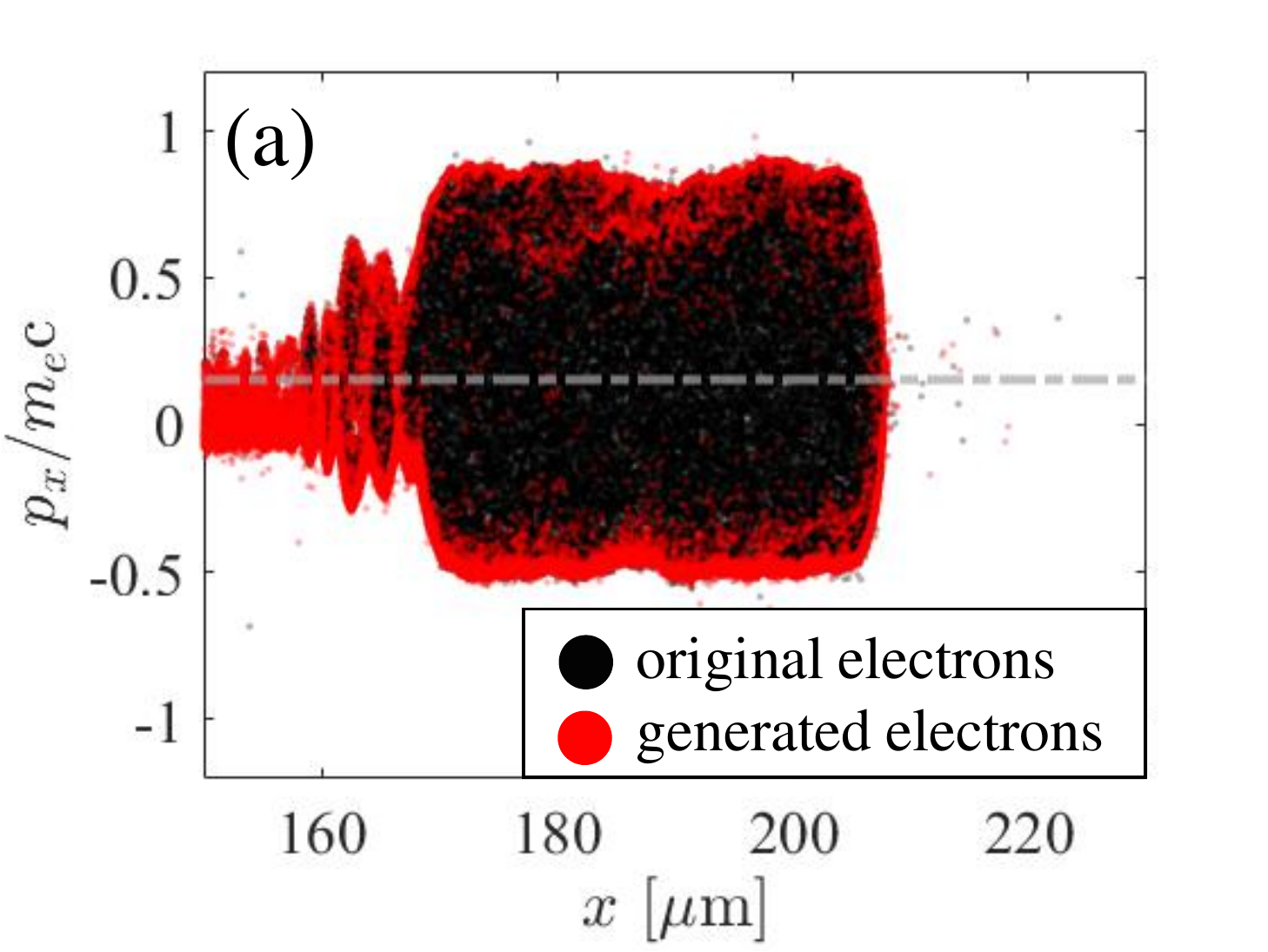}
\end{minipage}
\begin{minipage}[b]{0.391\textwidth}
\includegraphics[width=1\textwidth]{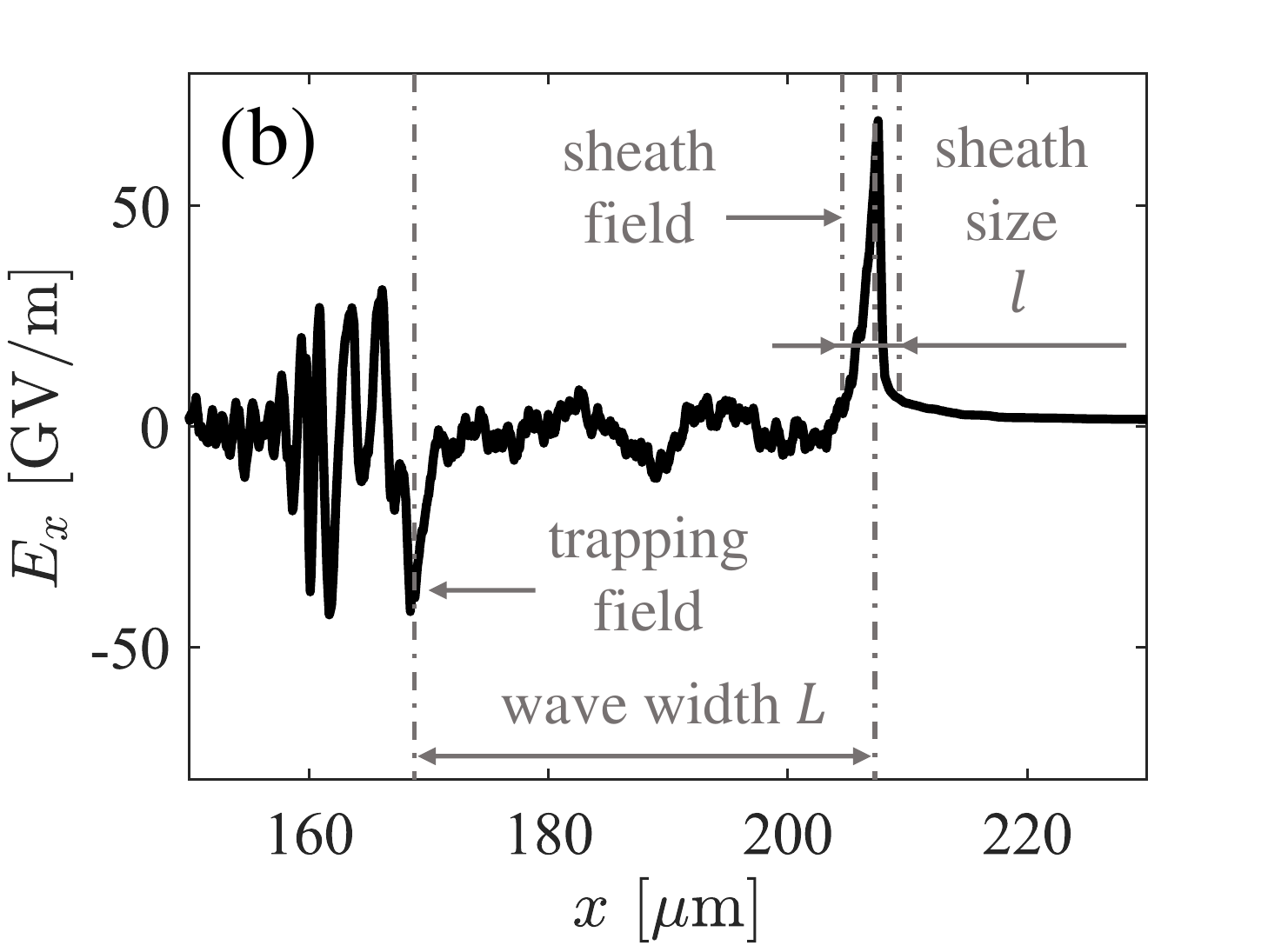}
\end{minipage}
\begin{minipage}[b]{0.391\textwidth}
\includegraphics[width=1\textwidth]{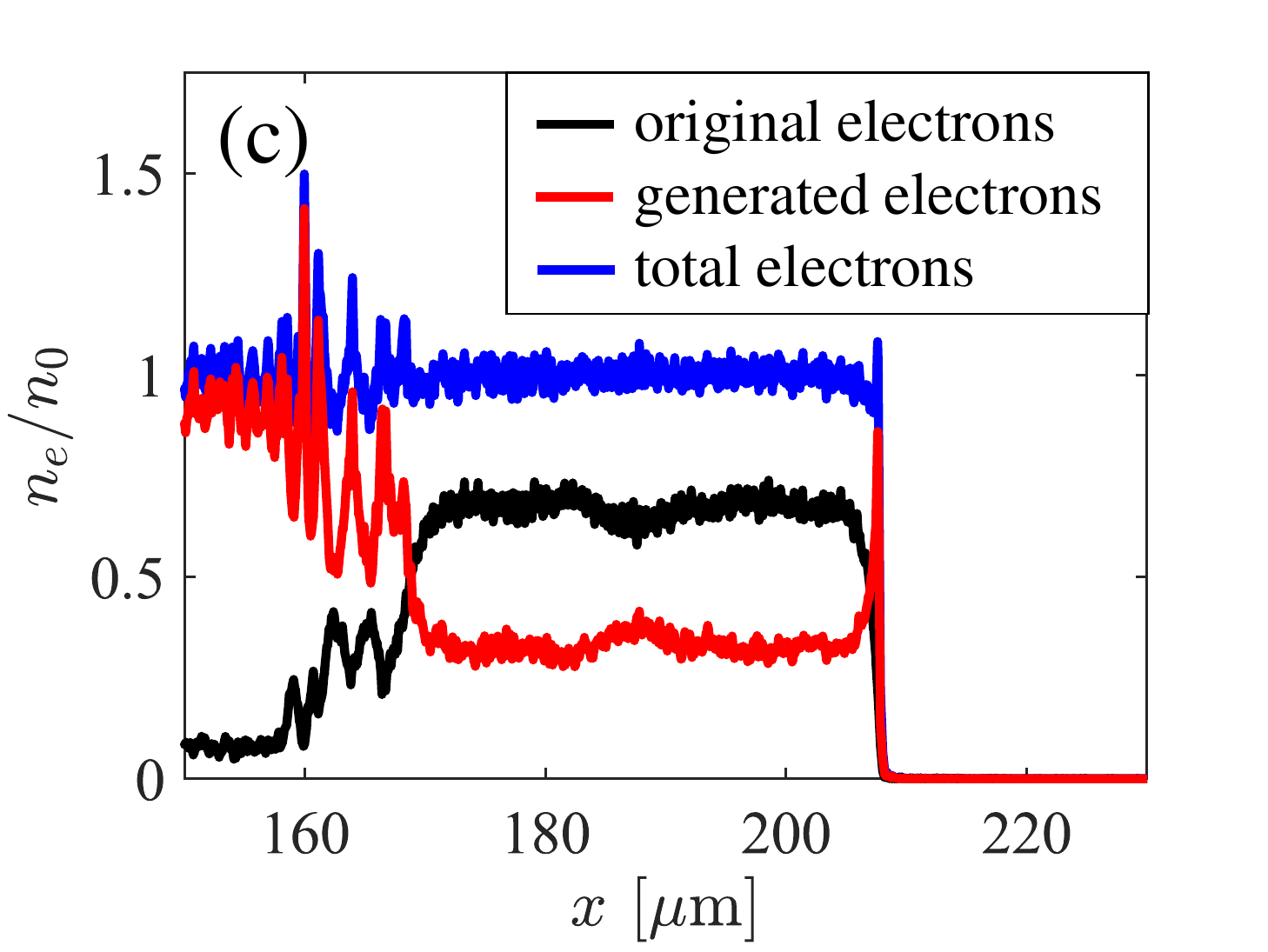}
\end{minipage}
\caption{Ionization wave structure without the applied magnetic field at $t = 2$~ps. (a)~Electron phase space. The grey dash-dotted line denotes the average momentum of the original electrons trapped by the wave. (b)~Longitudinal electric field $E_x$. (c)~Electron density, normalized by the initial electron density $n_0$.}
\label{fig1-ver2}
\end{figure}

As shown in Fig.~\ref{fig1-ver2}(b), the nonlinear ionization wave has two characteristic spatial scales: the width of the sheath field spike $l$ and the distance $L$ between this spike and the spike of the trapping field. The width of the sheath field is approximately equal to the Debye length, $\lambda_D$, calculated for $n_e \approx n_0$ and the characteristic energy $\varepsilon_{hot}$ of the hot electron population trapped by the wave:
\begin{equation} \label{eq:l}
    l \approx \lambda_D =  \sqrt{\varepsilon_{hot} /4 \pi n_0 e^2},
\end{equation}
where $e$ is the electron charge. The distance between the spikes can significantly exceed $l$, but a nonlinear analytical solution is required to determine the exact scaling for $L$. At this stage it is sufficient to point out that $L > l$. An applied magnetic field, $B_0$, introduces an additional characteristic spatial scale, which is the electron gyroradius:
\begin{equation} \label{eq:gyroradius}
    \rho_e \approx v_{hot} / \omega_{ce},
\end{equation}
where $v_{hot} \approx \sqrt{2 \varepsilon_{hot} / m_e}$ is the characteristic electron velocity (our assumption is that the hot electrons are weakly relativistic) and $\omega_{ce} =|e| B_0/m_e c$ is the electron gyrofrequency.

\begin{figure}
\centering
\begin{minipage}[b]{0.45\textwidth}
\includegraphics[width=1\textwidth]{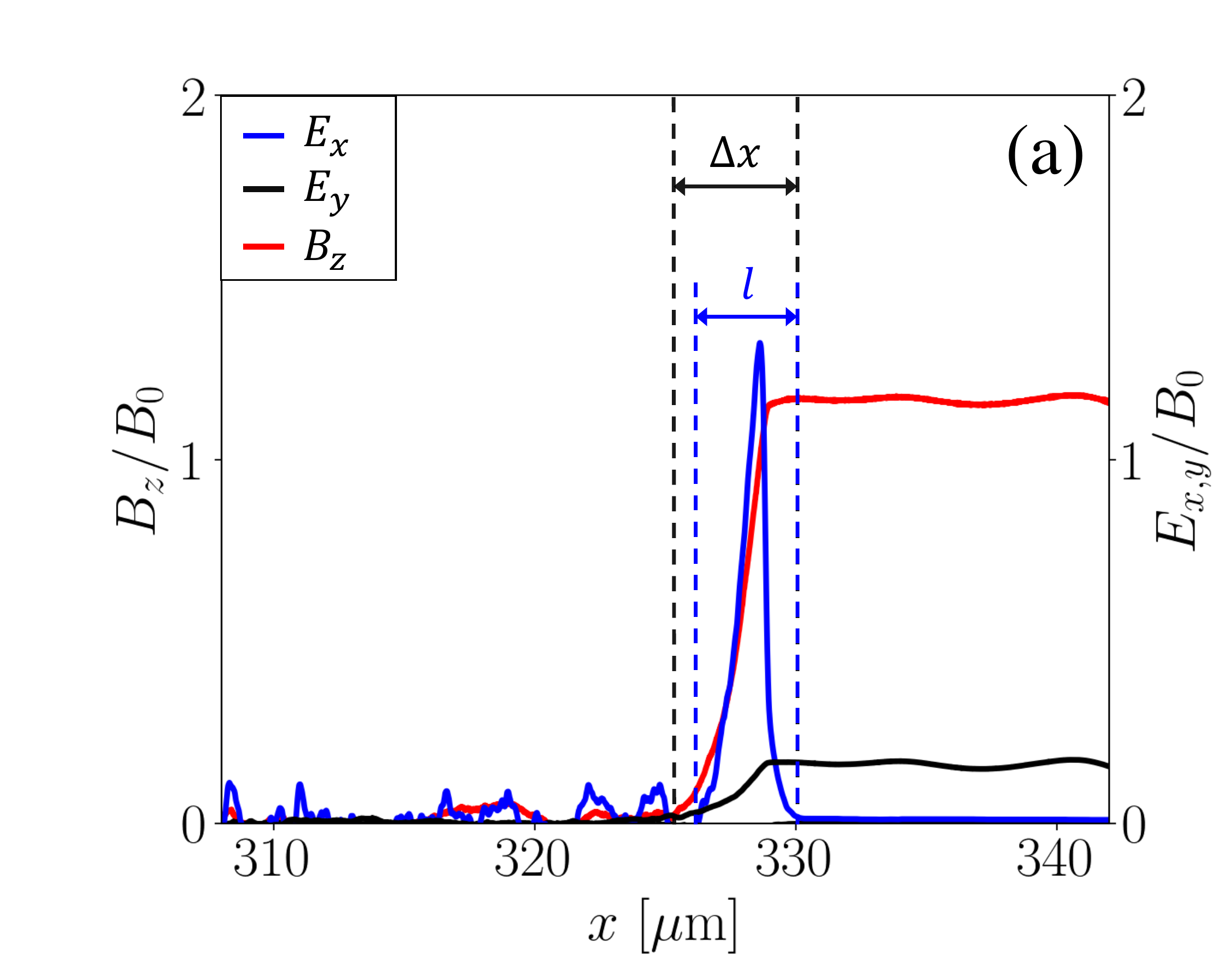}
\end{minipage}
\label{gradient-300(2)}
\begin{minipage}[b]{0.45\textwidth}
\includegraphics[width=1\textwidth]{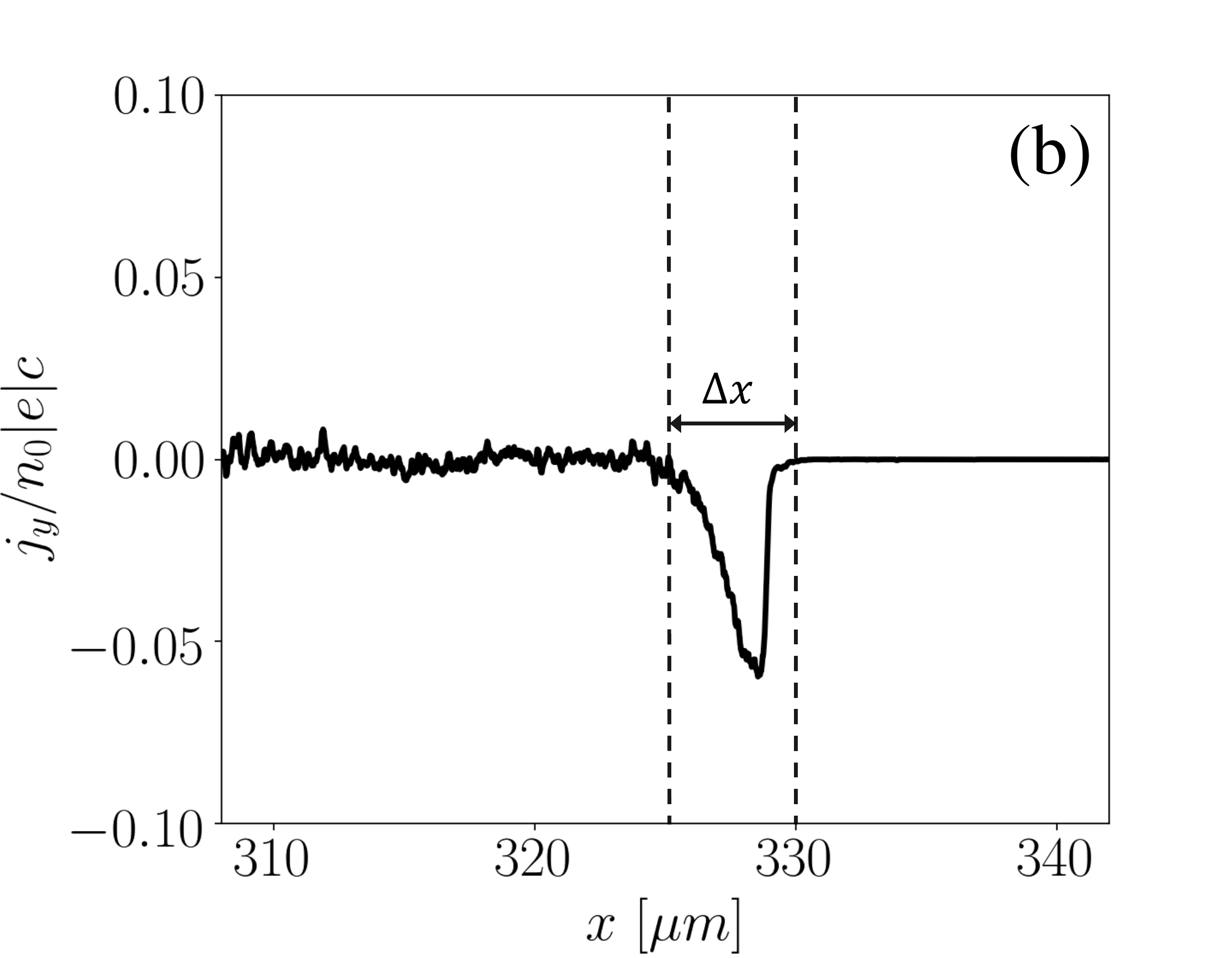}
\end{minipage}
\caption{Structure of the sheath region in the ionization wave propagating through neutral gas with an applied magnetic field ($B_0 = 2$~MG). (a)~Structure of electric and magnetic fields in the vicinity of the sheath region with thickness $l$. (b)~Structure of a current layer with thickness $\Delta x$ generated by the applied magnetic field. These snapshots are taken at $t=$5~ps.}
\label{fig:sheath current}
\end{figure}

As later confirmed by our simulations, the plasma is able to eliminate or shield out the magnetic field by generating a surface current layer, with the current flowing along the $y$-axis. Figure~\ref{fig:sheath current} provides a relevant example from a simulations whose details are discussed in the next section. The current is generated as the velocity of individual electrons is rotated by the magnetic field. The corresponding transverse velocity resulting from relatively weak rotation is $v_y \approx v_{hot} \omega_{ce} \Delta t$, where we have assumed that $v_y \ll v_{hot}$. Here $\Delta t \approx \Delta x / v_{hot}$ is the electron travel time through the current layer of thickness $\Delta x$, which is the only region with a magnetic field accessible to the plasma electrons. The field is fully shielded by the current layer if
\begin{equation} \label{eq:shielding}
    \frac{B_0}{\Delta x} \approx \frac{4 \pi}{c} |e| n_e v_y.
\end{equation}
We now take into account the estimates for $v_y$ and $\Delta t$ to find that
\begin{equation}
    v_y \approx \omega_{ce} \Delta x.
\end{equation}
We substitute this expression into Eq.~(\ref{eq:shielding}) to obtain an estimate for the thickness of the current layer:
\begin{equation} \label{eq:layer width}
    \Delta x \approx c / \omega_{pe},
\end{equation}
where $\omega_{pe} = \sqrt{4 \pi n_0 e^2 / m_e}$.

Our estimate is consistent with the assumption that the electrons perform only a relatively weak rotation if $\Delta x \ll \rho_e$. It is useful to introduce a new parameter, $\beta_0$, that is the ratio of the pressure of the hot electrons trapped by the ionization wave to the pressure of the applied magnetic field. It then follows from Eqs.~(\ref{eq:gyroradius}) and (\ref{eq:layer width}) that
\begin{equation}
    \Delta x \approx \rho_e  / \sqrt{\beta_0},
\end{equation}
where 
\begin{equation} \label{eq:beta0}
    \beta_0 \approx 8 \pi n_0 \varepsilon_{hot}/ B_0^2.
\end{equation}
We thus conclude that the applicability condition for our analysis is $\beta_0 \gg 1$.

To compare the thickness of the current layer with the width of the sheath, we use Eqs.~(\ref{eq:layer width}) and (\ref{eq:l}). We obtain that
\begin{equation}
    \Delta x / l \approx \sqrt{ m_e c^2 / \varepsilon_{hot} } .
\end{equation}
This ratio is independent of the magnetic field and it only depends on the characteristic energy of the electron population. In our regime of interest, the hot electrons are weakly relativistic, so the current layer is comparable to the sheath region and, more importantly, $\Delta x$ is much smaller than the width of the ionization wave $L$ (note that $L \gg l$ in our examples).

The parameters for the case without the magnetic field (see Sec.~\ref{sec:non-mag}) are derived from experimental conditions used to observe the ionization wave in the past~\cite{mccormick2014soliton}. It follows from Eq.~(\ref{eq:beta0}) that a 6~MG magnetic field is required to achieve $\beta_0 = 1$ at $\varepsilon_{hot} \approx 200$~keV. Such a strong field is beyond what is currently available, so we focus our study on somewhat lower magnetic fields in the MG-range that can potentially be accessed experimentally. Our estimates are valid for this regime of $\beta_0 > 1$, which indicates that we can expect for the plasma to shield out the magnetic field on a spatial scale that is smaller than the typical electron gyroradius. The implication of this is that the applied magnetic field introduces no major disruptions to the structure of the ionization wave and its primary manifestation is a thin current layer at the plasma edge. As the plasma boundary expands due to the ionization wave, the layer is expected to move outwards with the speed of the ionization wave.

%Note that the current layer is contained within the sheath region, $\Delta x \leq l$, for the considered electron distribution and $\beta_0 \gg 1$. For instance, at $\beta_0 \approx 10$, the corresponding condition on electron energy reads $\varepsilon_{hot} > m_e c^2 / \beta_0^2 \approx 5$~keV. In our case, the characteristic electron energy is in the range of 100~keV, so the condition is easily satisfied. 

We are now well positioned to examine the impact of the applied magnetic field on the propagation of the ionization wave. We established in Sec.~\ref{sec:non-mag} that the speed of the ionization wave changes slowly with time. This observation motivates us to consider a simple model where the current layer, $j_y$, is moving in the positive direction along the $x$-axis with a constant velocity equal to the velocity of the ionization wave and denoted as $u$. This current is only coupled to $E_y$ and $B_z$, so the corresponding wave equations read
\begin{eqnarray}
&& -\frac{\partial B_z}{\partial x} = \frac{4 \pi}{c} j_y + \frac{1}{c} \frac{\partial E_y}{\partial t}, \label{eq_1} \\
&& \frac{\partial E_y}{\partial x} = - \frac{1}{c} \frac{\partial B_z}{\partial t}. \label{eq_2}
\end{eqnarray}
We look for a solution that (in the vicinity of the current layer) depends on $x$ and $t$ through $s \equiv x - ut$. It follows from Eq.~(\ref{eq_2}) that 
\begin{eqnarray} \label{eq:2.2}
&& \frac{d E_y}{ds} = \frac{u}{c} \frac{d B_z}{ds}.
\end{eqnarray}
We have $E_y = B_z = 0$ to the left of the layer, i. e. inside the plasma. We use this boundary condition to integrate Eq.~(\ref{eq:2.2}) across the current layer, which yields
\begin{equation} \label{eq:E_y}
    E_y = \frac{u}{c} B_z.
\end{equation}
We next use this relation to re-write Eq.~(\ref{eq_1}) as
\begin{equation} \label{eq:j_y}
    \frac{d B_z}{ds} = - \frac{4 \pi}{c} j_y \left[ 1 - \frac{u^2}{c^2} \right]^{-1}.
\end{equation}

The electric field $E_y$ represents an electromagnetic wave emitted by the moving current layer. We make this explicit by representing the magnetic field outside of the plasma, i. e. in the ambient gas, as
\begin{equation}
    B_z = B_0 + B_1 = B_0 + E_y,
\end{equation}
where $B_0$ is the applied magnetic field and $B_1 = E_y$ is the amplitude of the electromagnetic wave. We use the expression for $E_y$ given by Eq.~({\ref{eq:E_y}) to find that the field structure in the surrounding gas consists of electric and magnetic fields with the following amplitudes:
\begin{eqnarray}
&& B_z = \frac{B_0}{1 - u/c}, \label{B_z estimate} \\
&& E_y = \frac{u}{c} \frac{B_0}{1 - u/c}, \label{E_y estiamte}
\end{eqnarray}
where the magnetic field is enhanced compared to $B_0$ due to the emissions of the electromagnetic wave.

The continuous emission of the electromagnetic radiation should lead to energy losses by the plasma electrons sustaining the ionization wave. The easiest way to find these losses is to integrate the product $(j_y E_y)$ over the entire current layer. We use Eqs.~(\ref{eq:E_y}) and (\ref{eq:j_y}) to express $E_y$ and $j_y$ in terms of $B_z$. As a result, we find that the rate of energy loss (per unit area of the plasma surface) is
\begin{eqnarray} \label{eq:el energy loss}
&&  \int j_y E_y dx = - u \left[ \frac{1+u/c}{1-u/c} \right] \frac{B_0^2}{8 \pi},
\end{eqnarray}
with the integration performed from inside the plasma, where $j_y = 0$, to the region occupied only by neutral gas, where again $j_y = 0$ [in this region $B_z$ is given by Eq.~(\ref{B_z estimate})].

We predict that the described energy loss causes the ionization wave to stop. We estimate the characteristic time $\tau$ that the wave travels before coming to a stop as the time that it takes to lose the majority of the electron kinetic energy possessed by the trapped electrons (per unit area of the plasma surface). The latter is approximately $\varepsilon_{hot} n_0 L$, so that the energy balance condition reads
\begin{equation}
    \frac{\varepsilon_{hot} n_0 L}{\tau} \approx u \left[ \frac{1+u/c}{1-u/c} \right] \frac{B_0^2}{8 \pi}.
\end{equation}
In the regime with $(1+u/c)/(1-u/c)\approx 1$, we have 
\begin{equation}
    \tau \approx \beta_0 L / u.
\end{equation}
The corresponding stopping distance, $l_{stop}$, is then independent of the wave velocity:
\begin{equation}
    l_{stop} \approx \beta_0 L.
    \label{estimation}
\end{equation}
The dependence on $\beta_0$ indicates that the stopping distance decreases with the magnetic field strength as $l_{stop} \propto 1 / B_0^2$. 

We can now compare the rate of energy loss by the electrons to the energy flux carried by the emitted electromagnetic wave. The $x$-component of the Poynting vector to the right of the current layer is 
\begin{equation}
    S_x = \frac{c}{4 \pi} E_y B_z = u \left[ 1 - u/c \right]^{-2}\frac{B_0^2}{4\pi}.
    \label{eq: poynting flux energy}
\end{equation}
In the limit of $1 - u/c \approx 1$, this is twice the rate of energy loss experienced by the electrons and given by Eq.~(\ref{eq:el energy loss}). The other half is the energy of the applied magnetic field eliminated by the moving current layer.

The key prediction of our qualitative analysis carried out in this section is that the magnetic field $B_0$ leads to energy losses associated with the emission of an electromagnetic wave. This effect should manifest itself as the ionization wave stopping over a distance that scales as $1 / B_0^2$.

% --------------------------------------------------------------------------

\section{Ionization wave propagation in a magnetic field} \label{sec:2MG}

In this section we present kinetic simulations that confirm qualitative predictions of Sec.~\ref{sec:estiamtes} for a regime with high $\beta_0$. 

An applied magnetic field can impact two aspects of the ionization wave physics -- the launch of the wave and its subsequent propagation. It is shown in Sec.~\ref{sec:non-mag} that, once the wave is launched, the electrons sustaining the wave become decoupled from the rest of the electron population. This aspect suggests that it is meaningful to study the two aspects (the launch and the propagation) separately. In order to make a quantitative comparison of the wave propagation with the case without the applied magnetic field, we consider a setup with a localized but uniform magnetic field in the region originally occupied by gas. Specifically, we set $B_z = B_0$ at $|x| \geq 200$~$\mu$m. This enables the HED plasma under consideration to launch the wave without any impact from the magnetic field. 

We have made changes to the 1D version of EPOCH (the PIC code used for this project) in order to implement this setup. The applied magnetic field is treated as a static external field that is added in the particle pusher to the magnetic field computed by the field solver. The field configuration implies that there is an external current $j^0_y$ at $|x| = 200$~$\mu$m that sustains the applied magnetic field. However, our implementation does not require us to involve this current into the calculations performed by the field solver to find the applied magnetic field. The current $j^0_y$ is not included into our simulations and we therefore ignore its energy exchange with the considered system. The justification for this is that $j^0_y E_y$ is only present for a very short time while the front of the ionization wave crosses the $|x| = 200$~$\mu$m plane.

Figure~\ref{B-200} that is discussed in more detail later in this section confirms that the ionization wave enters the magnetic field region without significant disruptions. All the changes associated with the magnetic field take place over a significant propagation distance while the wave is traveling through a region with a uniform magnetic field. The external current $j^0_y$ located at the initial magnetic boundary (if we were to included it into our simulations) is left behind the wave, so that its exact location and its effect on the wave become inconsequential. It is important to point out that our setup works well only in the high-$\beta$ regime. At $\beta_0$ close to unity, we observe significant disruptions to the electron population during the entrance of the wave into the magnetic field region and our setup loses its purpose.

\begin{figure}
\centering
\begin{minipage}[b]{0.45\textwidth}
\includegraphics[width=1\textwidth]{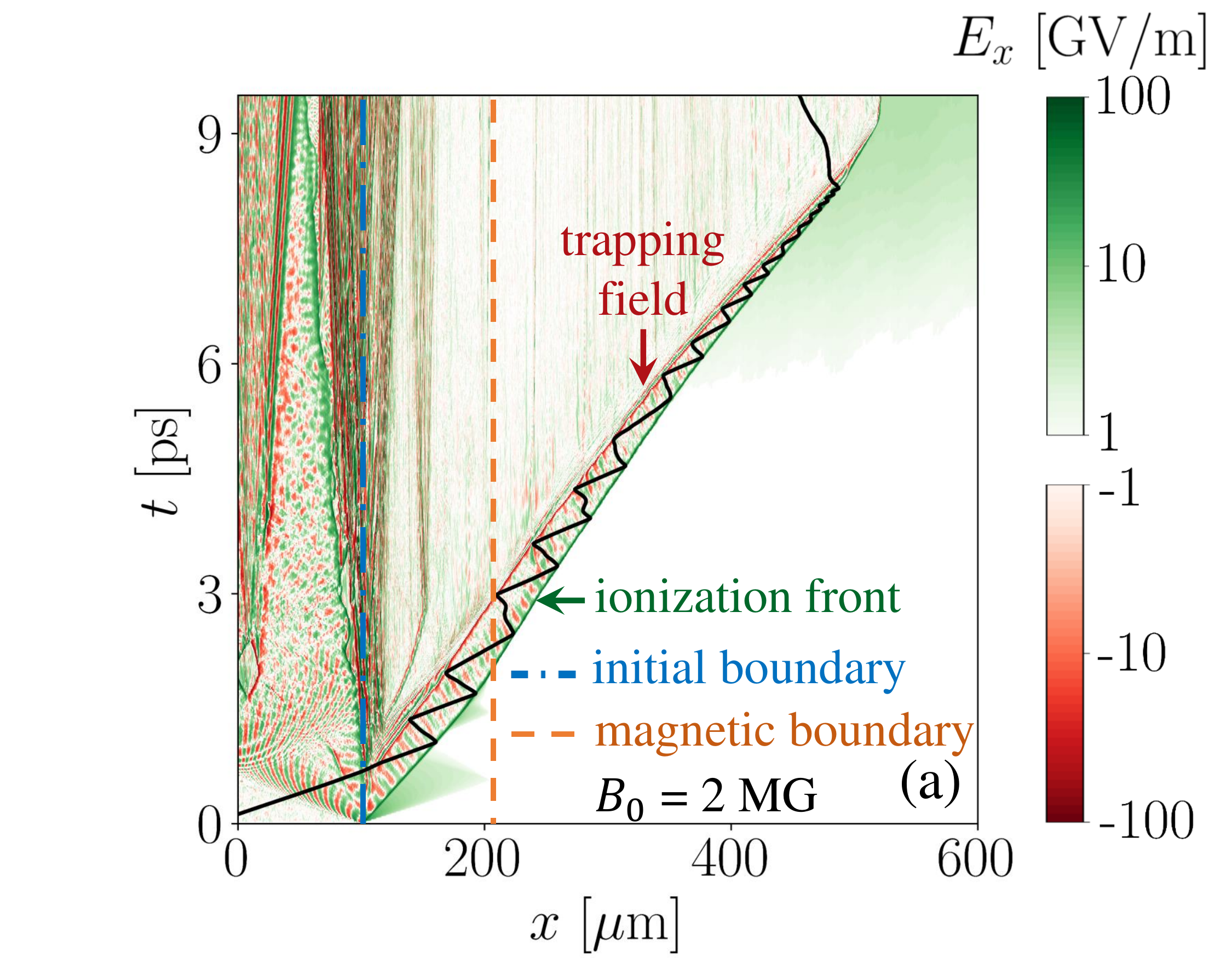}
\end{minipage}
\centering
\begin{minipage}[b]{0.45\textwidth}
\includegraphics[width=1\textwidth]{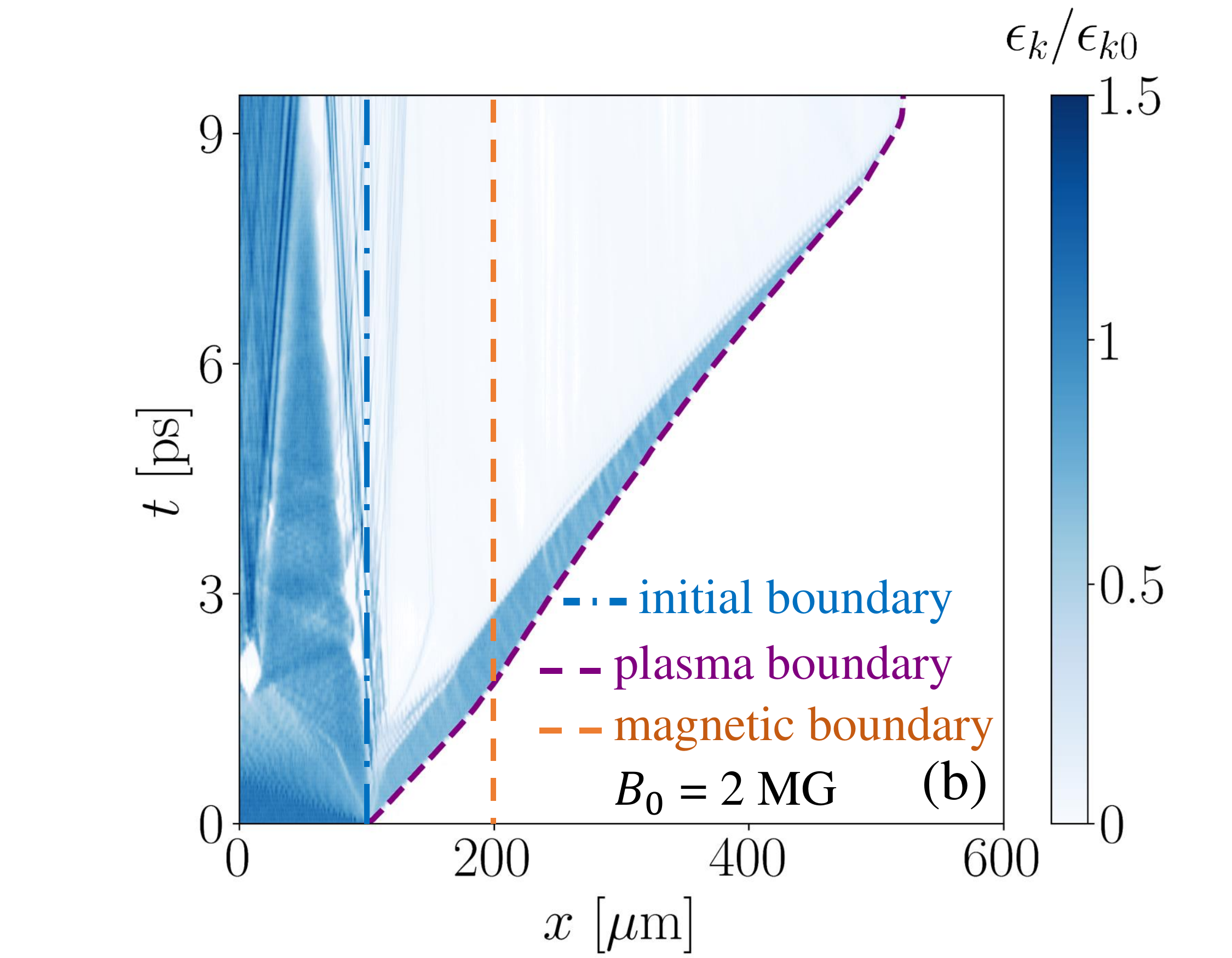}
\end{minipage}
\centering
\caption{Plasma expansion in neutral gas with an applied magnetic field of $B_0=2$~MG in $|x|>200~\mu$m. (a)~ Time evolution of $E_x$. The black curve shows a representative trajectory of a trapped energetic electron from the original population. (b)~Electron kinetic energy density $\epsilon_k$ normalized to its initial value $\epsilon_{k0}$. The ionization wave stops at 9~ps.
}
\label{B-200}
\end{figure}

Figure~\ref{fig:sheath current} shows the field and current layer structure in the wave that has travelled about 130~$\mu$m through the applied magnetic field with $B_0 = 2$~MG. The wave is launched by the HED plasma with parameters detailed in Table~\ref{simulation} (the same as in the case without the applied field considered in Sec.~\ref{sec:non-mag}). This is a high-$\beta$ regime with $\beta_0 \approx 9$. In good agreement with the estimates from Sec.~\ref{sec:estiamtes}, the magnetic field is shielded by the current layer whose thickness, $\Delta x$, is much smaller than wave width $L$ [see Fig.~\ref{fig1-ver2}(b)]. The layer emits an electromagnetic wave. We find that $E_y \approx 0.16~B_0$, which agrees well with Eq.~(\ref{E_y estiamte}) that predicts $E_y \approx 0.16~B_0$ for $u \approx 0.15~c$.

The enhancement of the magnetic field $B_z$ to the right of the sheath from $B_0$ to $1.16~B_0$ is a result of the electromagnetic wave emission. This enhancement is well captured by Eq.~(\ref{B_z estimate}). There is a wave-front associated with the electromagnetic wave (not shown) that propagates with the speed of light. Ahead of this wave-front, $E_y = 0$ and $B_z = B_0$. Our simulations use open boundary conditions for the fields, which means that the emitted wave leaves the simulation domain upon reaching the right boundary at $x = 600$~$\mu$m. The use of open boundary conditions prevents a magnetic field pile-up. It is worth pointing out that the strength of the magnetic field increases if we use reflecting boundary conditions. However, we assume that the considered system is large enough to ignore any impact from wave reflection on a time scale comparable to the time it takes the ionization wave to stop. 

Figure~\ref{B-200} shows the time evolution of $E_x$ and the electron kinetic energy density $\epsilon_k$ for the considered regime. As expected, the key features of the ionization wave are preserved. The wave maintains strong ionizing and trapping fields due to the electron trapping within the wave. We find that the propagation velocity remains roughly constant. The most noticeable change compared to the case without the applied field (see Fig.~\ref{B-0}) is the gradual reduction of the wave width. The wave structure becomes disrupted at about 9 ps, which stops the expansion of the plasma boundary. 

%%The distance that the wave has travelled in the presented simulation through the region initially occupied by the magnetic field is roughly 350~$\mu$m. The estimate given by Eq.~(\ref{estimation}) predicts $l_{stop} \approx $~$\mu$m for $\beta_0  = 8$ and $L = $~$\mu$m. This indicates that our estimate that accounts for the energy loss caused by the emission of the electromagnetic wave qualitatively captures the observed effect of the ionization wave stopping. 

%--------------------------------------------------------------------------
%--------------------------------------------------------------------------

\section{Stopping of the ionization wave in a magnetic field} \label{sec:stopping}

In Section~\ref{sec:2MG}, we presented simulation results for the case with $\beta_0 \approx 9$ that confirm the magnetic field expulsion by the ionization wave and the resulting emission of an electromagnetic wave. In this section, we examine the stopping of the ionization wave for a range of applied magnetic fields.

%by performing a $\beta_0$-scan by varying the strength of the applied field. 
 
Figure~\ref{beta} shows the distance travelled by the ionization wave before stopping as a function of $\beta_0$. The scan is performed by setting $B_0$ equal to 1.6, 1.7, 1.8, 1.9 and 2~MG and performing the corresponding PIC simulations (the plasma is set up with the same parameters as shown in Table~\ref{simulation}). The value of $\beta_0$ is determined using the kinetic energy density of the original electrons in the ionization wave right before it enters the applied magnetic field. The travel distance is determined from a plot similar to that shown in Fig.~\ref{B-200}(b) for $B_0 = 2$~MG that gives the time evolution of the plasma boundary for each simulation. We define $l_{stop}$ as the total displacement of the plasma boundary in the region initially occupied by the magnetic field. Based on our $\beta_0$ scan, we find that the stopping distance $l_{stop}$ increases linearly with $\beta_0$ as $l_{stop} = \kappa \beta_0 L$, where $\kappa \approx 0.9$ and $L$ is the width of the ionization wave right before it enters the magnetic field region. This is in good agreement with Eq.~(\ref{estimation}) of Sec.~\ref{sec:estiamtes}.

\begin{figure}
\centering
\begin{minipage}[b]{0.45\textwidth}
\includegraphics[width=1\textwidth]{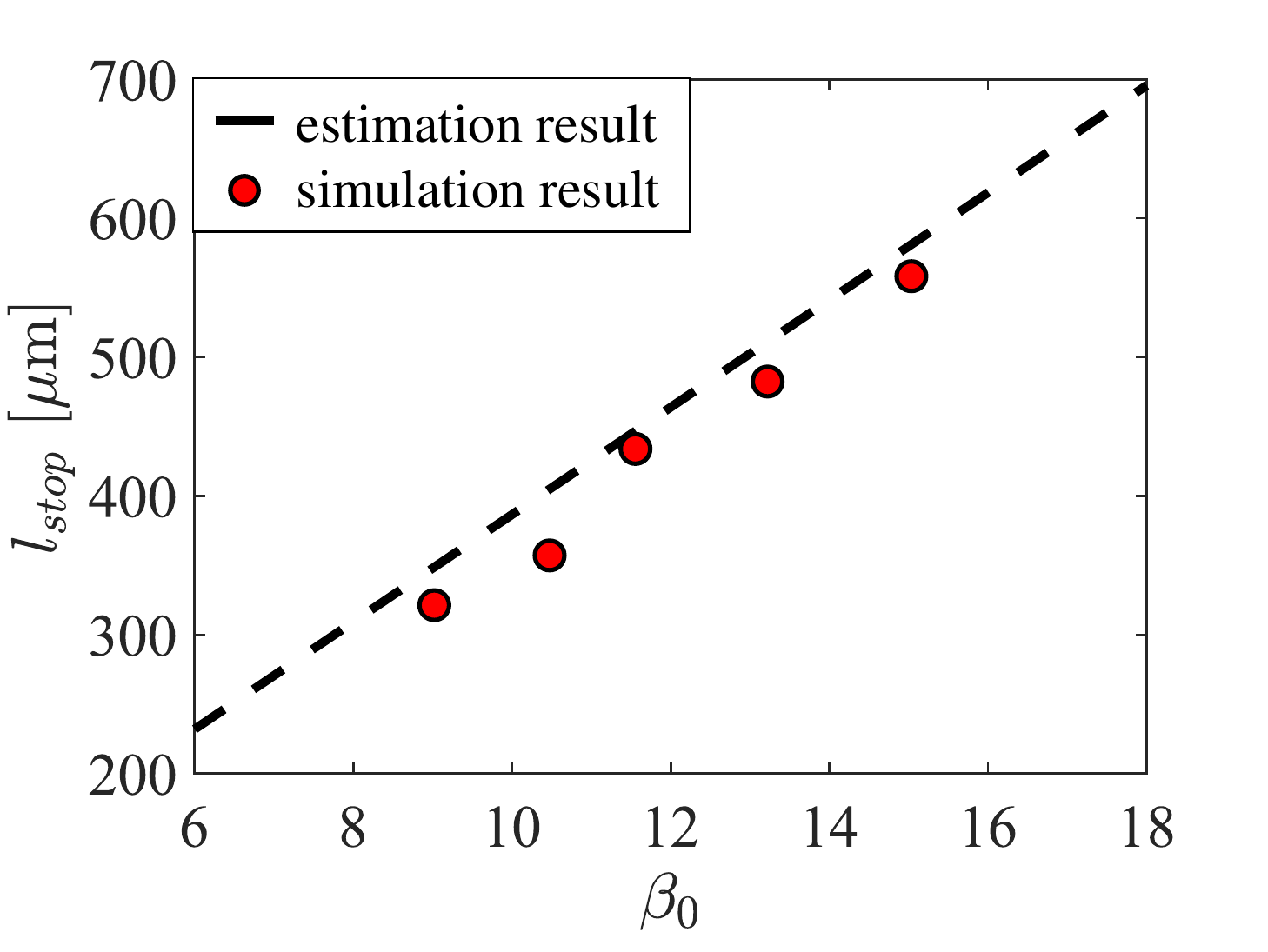}
\end{minipage}
\caption{Scan over the applied magnetic field strength $B_0$. Each marker represents a separate PIC simulation, with $l_{stop}$ being the total distance travelled by the ionization wave in the region with the applied magnetic field and $\beta_0$ is the ratio of the initial plasma pressure to that of the applied magnetic field. For all the cases presented, the emission of the electromagnetic wave accounts for $\sim 70\%$ of the energy initially associated with the ionization wave. %The color indicates the relative energy loss by the ionization wave associated with the emission of an electromagnetic wave [see Eq.~(\ref{eq:main energy balance}) for details].
The dashed line is $l_{stop} = \kappa\beta_0L$, where $\kappa= 0.9$. }
\label{beta}
\end{figure}

%%Simulation and estimation result of the travel distance of the ionization wave inside various applied magnetic field. The black bash line shows our estimation result as shown in Eq.~(\ref{estimation}). We add a correction to Eq.~(\ref{estimation}) by $l_{stop}\approx \kappa\beta_0L$, where $\kappa= 0.909$. The scatter shows the simulation result. $\beta_0 = 8\pi\varepsilon_{hot}n_0/B_0^2$ is the ratio of kinetic energy density of the hot electron inside the trapping wave and the magnetic field energy density of the applied magnetic field. Both $\beta_0$ and $L$ are measured at the time when the ionization wave hits the magnetic field boundary. The marker filled color shows the energy loss of electrons which goes to the electromagnetic wave.Detailed discussion of the energy balance is presented in Section~\ref{sec:stopping}

The ionization wave has two primary energy loss mechanisms: electromagnetic wave emissions and energetic electron losses associated with the electron leakage into the region behind the ionization wave. Using our simulation results, we are able to compare these two energy loss mechanisms and confirm that the wave emission is the dominant mechanism in the high-$\beta$ regime considered in this paper. We perform this analysis by splitting the simulation domain into three regions, as shown in Fig.~\ref{fig:energy balance}: region 1 is behind the ionization wave; region 2 is the ionization wave; region 3 is in front of the ionization wave. The boundaries separating these regions, $x_{min}$ and $x_{max}$, move with the ionization wave. The energy emitted by the electromagnetic wave is calculated by time-integrating the $x$-component of the Poynting vector at a fixed location, $x = x_b$. In what follows, we detail the calculations performed to quantify the two energy loss mechanisms.

%In our estimates presented in Sec.~\ref{sec:estiamtes}, we only accounted for the energy loss associated with the emission of the electromagnetic wave, but the ionization wave can also lose energy by losing energetic electrons. Using our simulation results, we are able to compare these two energy loss mechanisms. Figure~\ref{fig:sheath current} shows how the calculation is performed by splitting the domain into three regions: region 1 -- the region behind the ionization wave; region 2 -- the ionization wave; region 3 -- the region in front of the ionization wave. The boundaries separating these regions ($x_{min}$ and $x_{max}$) move with the ionization wave. The energy emitted by the electromagnetic wave is calculated by time-integrating the $x$-component of the Poynting vector at a fixed location, $x = x_b$.

We have verified that the energy is conserved in our simulations with good accuracy, so that, as the ionization wave moves forward through the region with the magnetic field, we have 
\begin{equation} \label{eq:energy conserv}
    {\cal{E}}_1 (t) + {\cal{E}}_2 (t) + {\cal{E}}_3 (t) + \int_{t_0}^t S_x (x_b) dt'  = I,
\end{equation}
where $I$ is a constant; ${\cal{E}}_1$, ${\cal{E}}_2$, and ${\cal{E}}_3$ are energies (per unit area) in regions 1, 2, and 3 at time $t$; 
\begin{equation}
    S_x (x_b) = \frac{c}{4 \pi} E_y (x_b) B_z (x_b)
\end{equation}
is the $x$-component of the Poynting vector at $x = x_b$. Here $t_0$ is the time of the ionization wave entering the magnetic field region. The energies in regions 1, 2, and 3 are calculated as 
\begin{eqnarray}
    {\cal{E}}_{1,2,3} = && \int_{x_L}^{x_R} \left[ \epsilon^e_k + \epsilon^i_k   + \frac{E^2 + B^2}{8\pi} \right] dx,
\end{eqnarray}
where $\epsilon^e_k$ and $\epsilon^i_k$ are the kinetic energy densities for electron and ions. The limits of integration are $x_L = 0$ and $x_R = x_{min}$ for region~1; $x_L = x_{min}$ and $x_R = x_{max}$ for region~2; and $x_L = x_{max}$ and $x_R = x_b$ for region~3. The energy loss by the ionization wave is then given by
\begin{eqnarray}
    {\cal{E}}_{init} - {\cal{E}}_2(t) &=& \left[ {\cal{E}}_1(t) - {\cal{E}}_1^0 \right] \nonumber \\
    &+& \left[ \int_{t_0}^t S_x (x_b) dt' + {\cal{E}}_3(t) - {\cal{E}}_3^0 \right], \label{eq:losses}
\end{eqnarray}
where ${\cal{E}}_1^0$, ${\cal{E}}_{init}$, and ${\cal{E}}_3^0$ are the energies in regions 1, 2, and 3 at $t = t_0$. Here we explicitly take into account that the electromagnetic wave emission starts at $t = t_0$, so that $I = {\cal{E}}_1^0 + {\cal{E}}_2^0 + {\cal{E}}_3^0$ at $t = t_0$. 

\begin{figure}
\centering
\includegraphics[width = 0.391\textwidth]{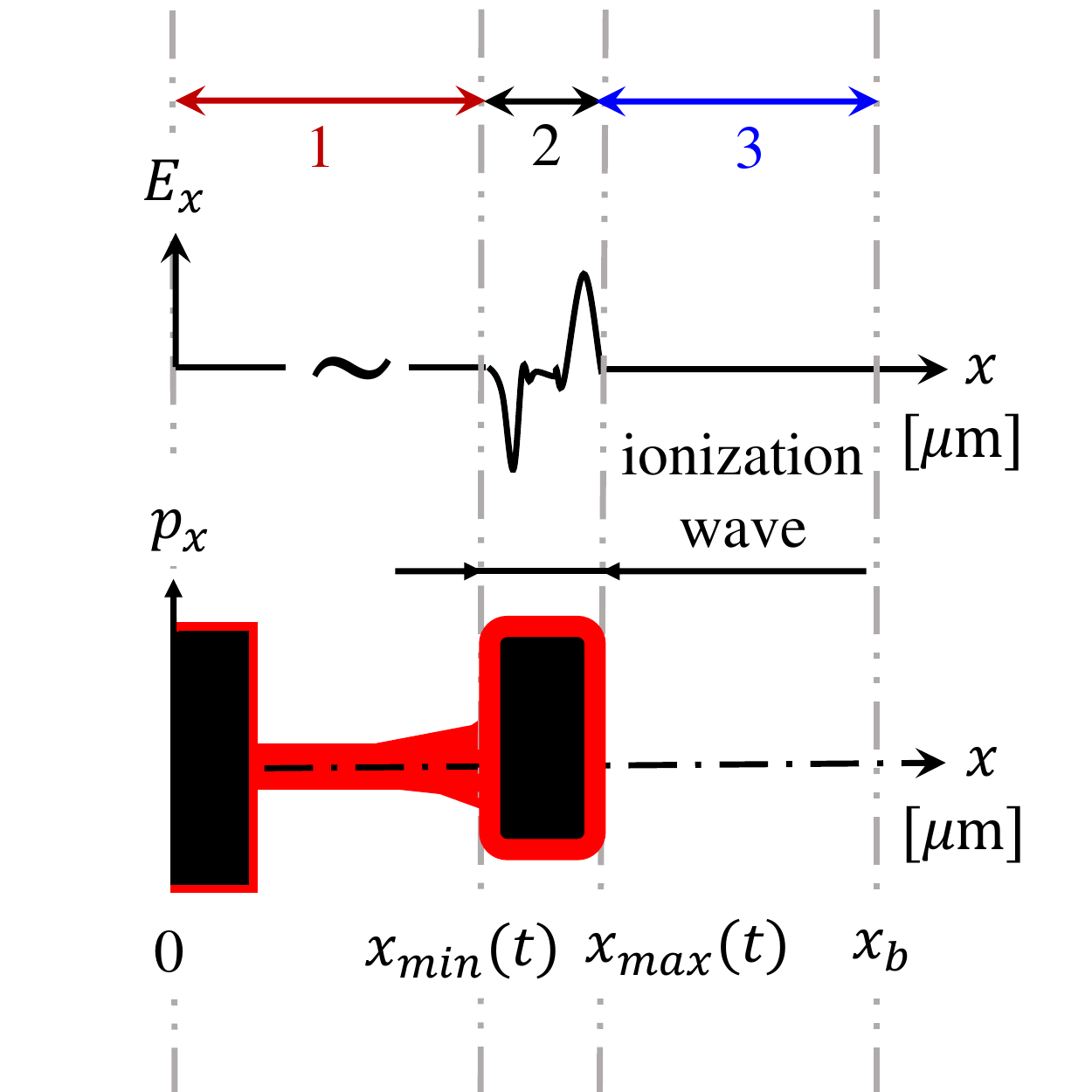}
\caption{Schematic setup for calculating the energy losses by the ionization wave. The front and the back of the ionization wave are located at $x_{max}(t)$ and $x_{min}(t)$. The energy emitted by the electromagnetic wave is calculated by integrating the Poynting flux through the plane located at $x = x_b$.}
\label{fig:energy balance}
\end{figure}

It is particularly convenient to analyze the energy loss that has taken place up to a given time $t$ in a setup where $x_{max}$ at this time $t$ is aligned with $x_b$. By definition, we have ${\cal{E}}_3 = 0$ when $x_{max} = x_b$. Moreover, ${\cal{E}}_3^0$ is the total energy of the applied magnetic field in the area swept by the ionization wave:
\begin{equation}
    {\cal{E}}_B \equiv {\cal{E}}_3^0 = \int_{x_{max}(t_0)}^{x_{max}(t) = x_b} \frac{B^2_0}{8\pi} dx,
\end{equation}
where $x_{max}(t_0)$ is the initial magnetic boundary in our simulation that is located at 200~$\mu$m. It then follows from Eq.~(\ref{eq:losses}) that the total energy loss ${\cal{E}}_{loss}$ by the ionization wave is  given by
\begin{equation} \label{eq:main energy balance}
    {\cal{E}}_{loss} \equiv {\cal{E}}_{init} - {\cal{E}}_2(t) = {\cal{E}}_{leak} + {\cal{E}}_{EM}  - {\cal{E}}_B, 
\end{equation}
where ${\cal{E}}_{leak} = {\cal{E}}_1(t) - {\cal{E}}_1^0$ is the total energy that has leaked into the region behind the ionization wave and
\begin{equation} \label{eq:energy conserv}
    {\cal{E}}_{EM} = \int_{t_0}^t S_x [x_{max} (t)] dt'
\end{equation}
is the total energy emitted by the electromagnetic wave.

Figure~\ref{Energy} shows the described analysis for the case with $B_0 = 2$~MG considered in Sec.~\ref{sec:2MG}. It is insightful to normalize all energies to ${\cal{E}}_{init}$, which is the energy in the ionization wave at $t = t_0$, i.e. right before it enters the magnetic field region. The vertical dash-dotted line indicates the time when the the plasma boundary stops expanding. We find the total energy emitted by the electromagnetic wave is approximately $1.2~{\cal{E}}_{init}$. Roughly 58\% of the energy comes from the ionization wave ($0.7~{\cal{E}}_{init}$), while the remaining 42\% of the energy comes from the energy of the displaced magnetic field ($0.5~{\cal{E}}_{init}$). This is in good agreement with our estimation in Eqs.~(\ref{eq:el energy loss}) and (\ref{eq: poynting flux energy}). In the context of the energy losses experienced by the ionization wave, our analysis indicates that the emission of the electromagnetic wave is the dominant mechanism that accounts for 70\% of the energy that is initially associated with the wave. 

%To conclude this section, we examine how the energy losses associated with the electromagnetic wave emission vary with $\beta_0$. The color in Fig.~\ref{beta} represents $({\cal{E}}_{EM}  - {\cal{E}}_B)/{\cal{E}}_{init}$. We find that this value changes insignificantly as we roughly double $\beta_0$ by increasing it from $\beta_0 \approx 9$ to $\beta_0 \approx 15$, with $({\cal{E}}_{EM}  - {\cal{E}}_B)/{\cal{E}}_{init} \approx 0.7$ for the range of considered $\beta_0$. This result indicates that the wave emission remains a dominant energy loss mechanism over the considered range of $\beta_0$. 
To conclude this section, we examine how the energy losses associated with the electromagnetic wave emission vary with $\beta_0$. The relative energy loss by the ionization wave associated with the emission of an electromagnetic wave can be expressed by $({\cal{E}}_{EM}  - {\cal{E}}_B)/{\cal{E}}_{init}$. We find that this value changes insignificantly as we roughly double $\beta_0$ by increasing it from $\beta_0 \approx 9$ to $\beta_0 \approx 15$, with $({\cal{E}}_{EM}  - {\cal{E}}_B)/{\cal{E}}_{init} \approx 0.7$ for the range of considered $\beta_0$. This result indicates that the wave emission remains a dominant energy loss mechanism over the considered range of $\beta_0$. 

\begin{figure}
\centering
\includegraphics[width=0.4\textwidth]{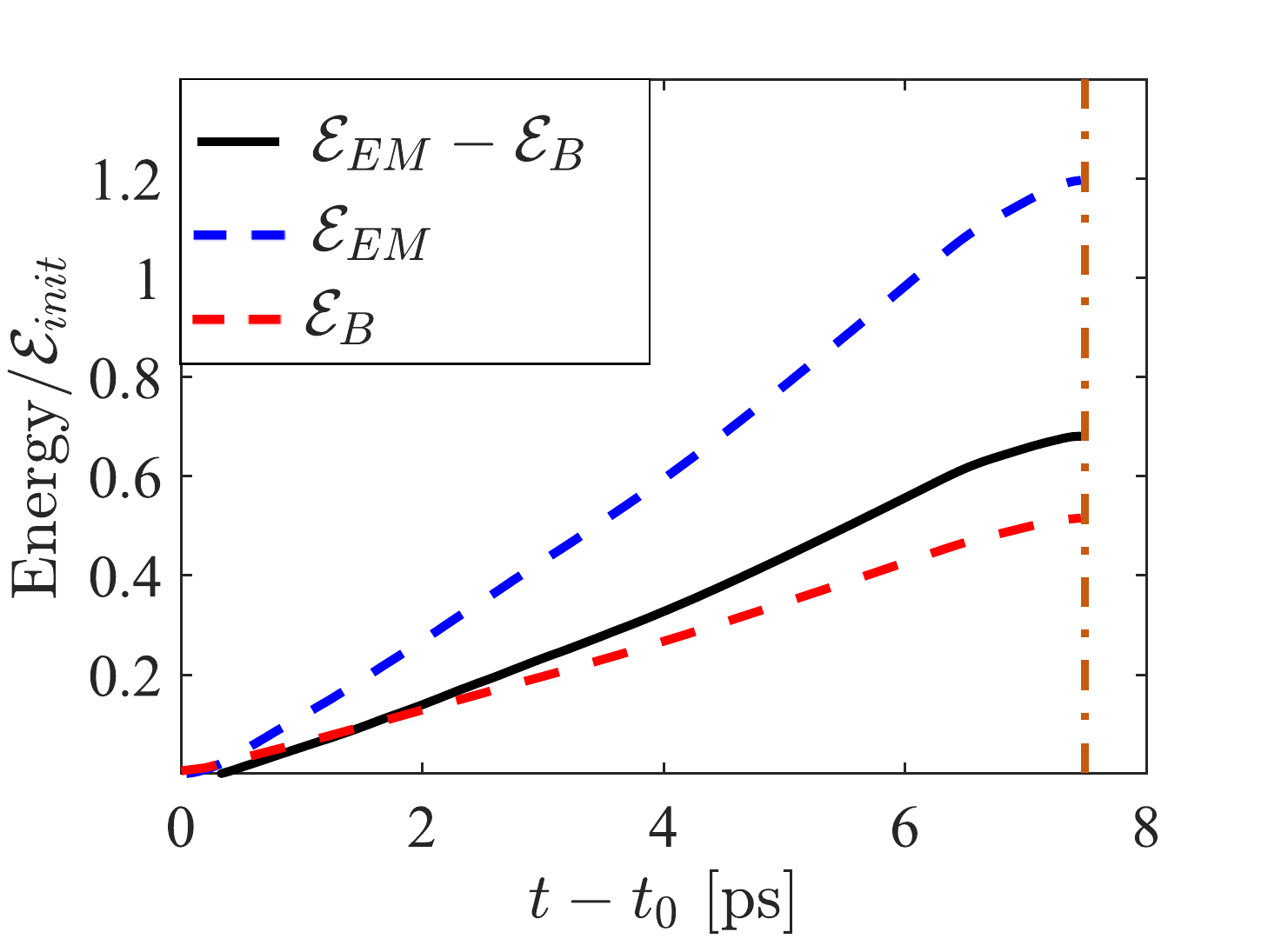}
\caption{Time evolution of the energy loss by the ionization wave associated with the electromagnetic wave emission (solid curve). The applied magnetic field is $B_0=$2~MG. The blue dashed (upper) curve is the total energy carried by the electromagnetic wave and the red dashed (lower) curve is the total energy of the expelled magnetic field [see Eq.~(\ref{eq:main energy balance}) for details]. The orange dash-dotted line denotes the time when ionization wave stops. The ionization wave enters the magnetic field region at $t = t_0$. The energies are normalized to ${\cal{E}}_{init}$, which is the energy in the ionization wave at $t = t_0$.}
\label{Energy}
\end{figure}

\section{Summary and Discussion} \label{sec:summary}

A high energy density plasma embedded in a neutral gas is able to launch an outward-propagating nonlinear electrostatic ionization wave that traps energetic electrons. The trapping maintains a strong sheath electric field, enabling rapid and long-lasting wave propagation aided by field ionization. In this paper, we have examined the propagation of a 1D ionization wave in the presence of a transverse MG-level magnetic field in a regime where the initial thermal pressure of the plasma exceeds the pressure of the magnetic field ($\beta > 1$). 

Our key finding is that the magnetic field stops the propagation by causing the energetic electrons sustaining the ionization wave to lose their energy by emitting an electromagnetic wave. The emission is accompanied by the magnetic field expulsion from the plasma. Our 1D3V kinetic simulations are supported by qualitative estimates for the stopping distance of the ionization wave [see Eq.~(\ref{estimation})] and an analytical emission model that predicts the amplitude of the electromagnetic wave based on the speed of the ionization wave and the amplitude of the applied magnetic field [see Eq.~(\ref{E_y estiamte})]. We find that the emission of the electromagnetic wave accounts for $\sim 70\%$ of the energy initially associated with the ionization wave. 

The described effect provides a mechanism mitigating rapid plasma expansion for those applications that involve an embedded plasma, such as high-flux neutron production from laser-irradiated deuterium gas jets. The examined magnetic field strength is in the range of what is now experimentally available~\cite{portugall1997field_gen,portugall1999field_gen,fujioka2013coil,santos2015coil,santos2018coil,gao2016coil,goyon2017coil,ivanov2018zebra}.

In order to perform a meaningful comparison with the regime without the applied magnetic field, we considered a setup where the wave is launched in the the region without the magnetic field. The wave enters a region with a uniform magnetic field only after it is fully formed. This configuration enabled us to develop a better understanding of the wave propagation and stopping in the presence of an applied magnetic field. In a potential experiment, the magnetic field is likely to surround the plasma from the very beginning. Therefore, a dedicated study that examines the launch of the ionization wave in the presence of an applied magnetic field would provide valuable insights.

The ionization wave formation and propagation has previously been demonstrated for an initially cylindrical plasma~\cite{mccormick2014soliton}. Such a plasma filaments during the expansion, producing expanding finger-like structures. Each filament maintains a high electron energy density, enabling long-lasting propagation analogous to the 1D case. A dedicated study based on the understanding developed in this work is needed to assess the impact of the magnetic field on the filamentation process.  

%+++++++++++++++++++++++++++++++++++++++++++++++++++++++++++++++++++++++++++

\begin{acknowledgments}
This research was supported by the DOE Office of Science under Grant No. DE-SC0019100. % put your number here instead
K. W. was supported by DOE Office of Science Grant No. DE-SC0018312.
Particle-in-cell simulations were performed using EPOCH~\cite{arber2015epoch}
developed under UK EPSRC Grant Nos. EP/G054940, EP/G055165, and EP/G056803.
This work used HPC resources of the Texas Advanced Computing Center (TACC) at the University of Texas at Austin.
Data collaboration was supported by the SeedMe2 project~\cite{chourasia2017seedme}
(http://dibbs.seedme.org).
\end{acknowledgments}

 %\nocite{*}
\providecommand{\noopsort}[1]{}\providecommand{\singleletter}[1]{#1}%
%

%\bibliography{aipsamp}% Produces the bibliography via BibTeX.

\end{document}